\documentclass[iop,revtex4,lscape]{emulateapj-rtx4}
\usepackage{multirow}
\usepackage{amsmath}
\usepackage{lscape}

\makeatletter

\newenvironment{inlinefigure}{%
\def\@captype{figure}%
\noindent\begin{minipage}{0.999\linewidth}\begin{center}}
{\end{center}\end{minipage}\smallskip}
\makeatother

\begin{document}

\title{The Host Galaxies of X-ray Quasars Are Not Strong Star Formers\altaffilmark{1,2,3,4,5}}

\author{
A.~J.~Barger\altaffilmark{6,7,8}, 
L.~L.~Cowie\altaffilmark{8},
F.~N.~Owen\altaffilmark{9},
C.-C.~Chen\altaffilmark{10},
G.~Hasinger\altaffilmark{8},
L.-Y.~Hsu\altaffilmark{8},
Y.~Li\altaffilmark{8}
}

\altaffiltext{1}{The James Clerk Maxwell Telescope has historically been operated by the Joint 
Astronomy Centre on behalf of the Science and Technology 
Facilities Council of the United Kingdom, the National Research 
Council of Canada and the Netherlands Organisation
for Scientific Research. 
Additional funds for the construction of SCUBA-2 
were provided by the Canada Foundation for Innovation.}
\altaffiltext{2}{The National Radio Astronomy Observatory is a facility of 
the National Science Foundation operated under cooperative agreement by 
Associated Universities, Inc.}
\altaffiltext{3}{The Submillimeter Array is a joint project between the
Smithsonian Astrophysical Observatory and the Academia Sinica Institute
of Astronomy and Astrophysics and is funded by the Smithsonian Institution
and the Academia Sinica.}
\altaffiltext{4}{The W.~M.~Keck Observatory is operated as a scientific
partnership among the California Institute of Technology, the University
of California, and NASA, and was made possible by the generous financial
support of the W.~M.~Keck Foundation.}
\altaffiltext{5}{{\em Herschel\/} is an ESA space observatory with science instruments provided
by European-led Principal Investigator consortia and with important participation from NASA.}
\altaffiltext{6}{Department of Astronomy, University of Wisconsin-Madison,
475 N. Charter Street, Madison, WI 53706, USA}
\altaffiltext{7}{Department of Physics and Astronomy, University of Hawaii,
2505 Correa Road, Honolulu, HI 96822, USA}
\altaffiltext{8}{Institute for Astronomy, University of Hawaii,
2680 Woodlawn Drive, Honolulu, HI 96822, USA}
\altaffiltext{9}{National Radio Astronomy Observatory, P.O. Box O, 
Socorro, NM 87801, USA}
\altaffiltext{10}{Institute for Computational Cosmology, Department of Physics, 
Durham University, South Road, Durham DH1 3LE, UK}

\slugcomment{In press at The Astrophysical Journal}


\begin{abstract}

We use ultradeep SCUBA-2 850~$\mu$m observations ($\sim0.37$~mJy rms)
of the 2~Ms {\em Chandra\/} Deep Field-North (CDF-N) and 4~Ms 
{\em Chandra\/} Deep Field-South (CDF-S) X-ray fields to examine the 
amount of dusty star formation taking place in the host galaxies of high-redshift X-ray 
active galactic nuclei (AGNs).
Supplementing with COSMOS, we measure the submillimeter fluxes of the $4-8$~keV 
sources at $z>1$, finding little flux at the 
highest X-ray luminosities but significant flux at intermediate luminosities.
We determine gray body and mid-infrared (MIR) luminosities by fitting spectral
energy distributions to each 
X-ray source and to each radio source in an ultradeep 
Karl G. Jansky Very Large Array (VLA) 1.4~GHz (11.5~$\mu$Jy at $5\sigma$) image of the CDF-N.
We confirm the far-infrared (FIR)-radio and MIR-radio correlations to $z=4$ using the 
non-X-ray detected radio sources. Both correlations are also
obeyed by the X-ray less luminous AGNs but not by the X-ray quasars.
We interpret the low FIR luminosities relative to the MIR for the X-ray quasars
as being due to a lack of star formation, while the MIR stays high due to the AGN contribution.
We find that the FIR luminosity distributions are highly skewed and the means are dominated
by a small number of high-luminosity galaxies. Thus, stacking or averaging analyses
will overestimate the level of star formation taking place in the bulk of the X-ray sample.
We conclude that most of the host galaxies of X-ray quasars are not strong star formers,
perhaps because their star formation is suppressed by AGN feedback.

\end{abstract}



\section{Introduction}
\label{intro}

A major open question in galaxy evolution is the interplay between star formation and active galactic nucleus (AGN) activity. 
Theoretical work has shown that ``feedback'' from an AGN can limit galaxy masses and luminosities by suppressing 
star formation, either through a powerful wind that clears the interstellar medium from the host galaxy (quasar-mode), 
or through the production of jets of relativistic particles that prevent gas in the hot halo from cooling (radio-mode)
(e.g., Ostriker \& Cowie 1981; Silk \& Rees 1998; Granato et al.\ 2004; Di Matteo et al.\ 2005; Springel et al.\ 2005; 
Bower et al.\ 2006; Croton et al.\ 2006; Hopkins et al.\ 2006; Sijacki et al.\ 2007).
Recently, observational evidence for the curtailing of star formation by radiatively driven outflows from AGNs 
has also been reported (e.g., Cano-D{\'i}az et al.\ 2012; Farrah et al.\ 2012).

The advent of sensitive, large-area, far-infrared (FIR) and submillimeter surveys from the
PACS (Poglitsch et al.\ 2010) and SPIRE (Griffin et al.\ 2010) instruments on the
ESA {\em Herschel Space Observatory\/} (Pilbratt et al.\ 2010) 
and the SCUBA-2 camera (Holland et al.\ 2013) on the 15~m James Clerk Maxwell Telescope (JCMT), 
along with deep X-ray observations from the NASA {\em Chandra X-ray Observatory\/}
(Weisskopf et al.\ 2002), 
has opened up a new avenue for exploring 
how AGNs can impact star formation in their host galaxies.
The mid-infrared (MIR; $5-40~\mu$m) fluxes of AGN hosts are dominated by thermal emission from 
hot dust, heated due to irradiation by the AGNs 
(e.g., Horst et al.\ 2008; Gandhi et al.\ 2009; Ichikawa et al.\ 2012). 
In contrast, it has recently been argued, based on the mapping of FIR wavelengths through the 
peak of the cold dust emission at 100~$\mu$m, that the FIR fluxes of AGN hosts 
are dominated by star formation (e.g., Hatziminaoglou et al.\ 2010; Mullaney et al.\ 2012).
This has led to studies with {\em Herschel\/} of the average star formation 
rates (SFRs) of AGN hosts selected from hard X-ray ($2-8$~keV) samples. 

Using the {\em Chandra\/} Deep Field-North (CDF-N) X-ray sample (Alexander et al.\ 2003) with
spectroscopic redshifts (Barger et al.\ 2008; Trouille et al.\ 2008) and
{\em Herschel\/} Multi-tiered Extragalactic Survey (HerMES) (P.I. S. Oliver; described in Oliver et al.\ 2012)
250, 350, and 500~$\mu$m imaging,
Page et al.\ (2012) found a systematic non-detection at 250~$\mu$m of the 21 $1<z<3$ AGN hosts with
X-ray luminosities $L_{2-8~{\rm keV}}>10^{44}$~erg~s$^{-1}$ (quasar luminosities; 
e.g., Barger et al.\ 2005; Richards et al.\ 2005). 

Page et al.\ (2012) also performed a stacking analysis of all the $1<z<3$ AGN hosts with X-ray luminosities
in a given range, whether they were detected at 250~$\mu$m or not, to probe below the confusion limit of the 
{\em Herschel\/} data. They derived an average SFR
for the $1<z<3$ AGN hosts with $L_{2-8~{\rm keV}}>10^{44}$~erg~s$^{-1}$ and found it to be considerably
lower than the average SFR
for the AGN hosts with $L_{2-8~{\rm keV}}=10^{43}-10^{44}$~erg~s$^{-1}$.

The Page et al.\ (2012) results may indicate that luminous AGNs are suppressing star formation, 
as would be expected from quasar-mode feedback. 
However, they are in contradiction with other {\em Herschel\/} stacking analyses, which find
average SFRs that either rise or stay flat to the highest X-ray luminosities
(e.g., Lutz et al.\ 2010; Shao et al.\ 2010; Rosario et al.\ 2012; Rovilos et al.\ 2012).

To try and resolve this discrepancy, Harrison et al.\ (2012) performed their own stacking
analyses on the 250~$\mu$m HerMES data in three fields: the CDF-N, the 
{\em Chandra\/} Deep Field-South (CDF-S), and COSMOS.
In their highest $L_X$ bin in the CDF-N, they confirmed Page et al.\ (2012)'s non-detection.
However, they postulated that low number statistics (they had seven sources in this bin) could 
be a problem. Thus, they analyzed
the wider area COSMOS data to improve the statistics. This time they
found constant average SFRs over the X-ray luminosity range 
$L_X=10^{43}-10^{45}$~erg~s$^{-1}$.
Their CDF-S results were within $1\sigma$ of their COSMOS results, while their 
CDF-N results----this time they used the GOODS-{\em Herschel\/}
(GOODS-H; P.I. D. Elbaz; described in Elbaz et al.\ 2011) data---were
low by $3\sigma$ compared to their COSMOS results.

A stacking analysis is a useful but necessarily blunt tool that hides a lot of information, 
since all one can get from a stacking analysis is an average.
If possible, it is much better to look at the spread in a quantity for
individual galaxies in order to determine what is happening. 
To do so, one needs exceptionally high quality data, both in the FIR/submillimeter
and in the X-ray. This requires the use of the deepest fields available.

In this paper, we use ultradeep SCUBA-2 observations of the CDF-N and CDF-S 
from Barger et al.\ (2014) and L. Cowie et al.\ (2015, in preparation)
to examine the amount of dusty star formation taking place
in the host galaxies of high-redshift X-ray AGNs.
These fields have incredibly deep X-ray data from Alexander et al.\ (2003; CDF-N; 2~Ms) and
Xue et al.\ (2011; CDF-S; 4~Ms).
In the first part of our analysis (Section~\ref{SF}), we supplement our primary fields
of study with the SCUBA-2 image 
of the central region of the COSMOS field from Casey et al.\ (2013), which has deep X-ray data from 
Elvis et al.\ (2009; C-COSMOS, 160~ks).

SCUBA-2's long-wavelength angular resolution on the sky is substantially 
better than that of space-based missions.  For example, 
the beam FWHM size of {\em Herschel\/} at its longest wavelength of 500~$\mu$m is $\sim35''$,
while that of SCUBA-2 at 850~$\mu$m is $\sim14''$.  
Previous work on this topic was primarily done using
{\em Herschel\/} at 250~$\mu$m, where the beam FWHM size is $\sim18''$. However,
the one source per 40 beams confusion noise in the {\em Herschel\/} data is (19, 18, 16)~mJy
at (250, 350, 500)~$\mu$m (Nguyen et al.\ 2010), compared to 2.1~mJy at 850~$\mu$m, with
the confusion being more dominated by low-redshift sources at the shorter wavelengths
(see, e.g., Fig.~10 of Casey et al.\ 2012).
Finally, for the redshift range $z=1$ to 5,
850~$\mu$m samples rest-frame wavelengths from 425~$\mu$m to 142~$\mu$m, while 250~$\mu$m
samples rest-frame wavelengths from $125~\mu$m to 42~$\mu$m, which pushes into the
MIR portion of the spectral energy distribution (SED) where the AGN torus is 
beginning to contribute to the emission.

In Section~\ref{data}, we describe the ultradeep X-ray
and radio samples (the latter covers only the CDF-N) that we use, along with the
corresponding optical, near-infrared (NIR), MIR, FIR, submillimeter, and millimeter 
imaging and optical spectroscopy.
In Section~\ref{SF}, we measure the submillimeter fluxes of
the X-ray sources in the CDF-N, CDF-S, and COSMOS fields and
find a significant dependence on X-ray luminosity (i.e., there is less submillimeter light in 
the most X-ray luminous AGNs).
In Section~\ref{SED}, we first construct the average SEDs of the X-ray sources
in the CDF-N and CDF-S fields to show schematically how the observed dependence on X-ray 
luminosity from Section~\ref{SF} is reflected in the FIR shapes. 
We then fit the SEDs of each CDF-N and CDF-S X-ray source and each CDF-N radio 
source individually at wavelengths longer than a rest-frame 
wavelength of $4~\mu$m with a combined gray body and truncated MIR power law.
In Section~\ref{xrayFIR-radio}, we use the resulting gray body luminosities to
confirm that the FIR-radio correlation holds to high
redshifts for the non-X-ray detected radio sample. We also determine that the X-ray quasars 
fall below the correlation, while the X-ray less luminous AGNs obey it.
In Section~\ref{xrayFIR-MIR}, we use the resulting MIR luminosities to confirm that the 
MIR-radio correlation holds to high redshifts for the non-X-ray detected radio sample.
We also determine that the X-ray quasars fall below the correlation, while the X-ray less 
luminous AGNs obey it. In Section~\ref{lumdist}, we analyze the FIR luminosity distribution
as a function of X-ray luminosity.
In Section~\ref{summary}, we summarize our results.

We adopt the AB magnitude system for the optical and NIR
photometry, and we assume the Wilkinson Microwave
Anisotropy Probe cosmology of $H_0=70.5$~km~s$^{-1}$~Mpc$^{-1}$,
$\Omega_{\rm M}=0.27$, and $\Omega_\Lambda=0.73$ 
(Larson et al.\ 2011) throughout.

\section{Data}
\label{data}

\subsection{X-ray Imaging}
\label{xray}

In order to provide a uniform sample, we choose AGNs solely on the basis of their hard
X-ray luminosities. 
In contrast, some papers in the literature (e.g., Shao et al.\ 2010; Mullaney et al.\ 2012) 
adopt the Bauer et al.\ (2004) mixed criteria (namely, X-ray luminosity, X-ray obscuring column 
or hardness, optical spectroscopic classifications, and X-ray/optical flux ratios) for separating 
X-ray AGNs from star formation dominated sources, sometimes in combination with a
{\em Spitzer Space Telescope\/} (Soifer et al.\ 2008) 
IRAC (Fazio et al.\ 2004) color-color selection (e.g., Chen et al.\ 2013).
However, with our pure hard X-ray luminosity, or, equivalently, 
black hole accretion rate selection,
we ensure that we will be comparing our measured FIR luminosities 
with genuine AGN luminosities. It will also simplify future comparisons with the 
{\em Swift\/}/Burst Alert Telescope local sample of Mushotzky et al.\ (2014).

To minimize opacity effects, we use the hardest {\em Chandra\/} X-ray band 
available ($4-8$~keV), which corresponds to a rest-frame energy selection of $8-16$~keV or 
greater for $z>1$. At these X-ray energies, opacity effects should be negligible, except for 
extremely Compton-thick sources ($N_H>10^{24}$~cm$^{-2}$). 
For the CDF-N, we start with the observed-frame $4-8$~keV sample of Alexander et al.\ (2003),
and for the CDF-S, we start with the observed-frame $4-8$~keV sample of Lehmer et al.\ (2012; 
catalog kindly supplied by B. Lehmer), taking the X-ray properties from Xue et al.\ (2011). 
Near the aim point, the X-ray data reach limiting fluxes of
$f_{\rm 4-8~keV}\approx 2\times 10^{-16}$~erg~cm$^{-2}$~s$^{-1}$ in the CDF-N
and $f_{\rm 4-8~keV}\approx10^{-16}$~erg~cm$^{-2}$~s$^{-1}$ in the CDF-S. 
For the COSMOS field, we generated a $4-8$~keV sample down
to $f_{\rm 4-8~keV}\approx 1.5\times 10^{-15}$~erg~cm$^{-2}$~s$^{-1}$
over the region that contains the deep SCUBA-2 data of Casey et al.\ (2013). 
This sample contains 96 sources.

\subsection{NIR and Optical Imaging}
\label{opt}

In the NIR and optical, we use the {\em Spitzer\/} IRAC 3.6~$\mu$m, 4.5~$\mu$m, 
5.8~$\mu$m, and 8.0~$\mu$m images (P.I. M. Dickinson), the
Canada-France-Hawaii Telescope
WIRCAM $K_s$-band images of Wang et al.\ (2010; CDF-N) and Hsieh et al.\ (2012; CDF-S)
deepened with additional data obtained after these papers were published,
the {\em Hubble Space Telescope (HST)\/} CANDELS images of Grogin et al.\ (2011)
and Koekemoer et al.\ (2011), the {\em Hubble Space Telescope (HST)\/} 
GOODS images of Giavalisco et al.\ (2004), 
the Subaru Suprime-Cam (Miyazaki et al.\ 2002) images of Capak et al.\ (2004; CDF-N), the
VLT images of Nonino et al.\ (2009; CDF-S), and the deep
{\em GALEX\/} (Martin et al.\ 2005) near-ultraviolet and far-ultraviolet images.

We astrometrically aligned the $K_s$-band images to the 
Karl G. Jansky Very Large Array (VLA) 1.4~GHz catalogs of F. Owen (2015,
in preparation; CDF-N) and Miller et al.\ (2013; CDF-S), which provide our absolute coordinate system.
Based on a subsample of radio sources with bright ($18.5-19.5$) $K_s$-band counterparts, 
the mean or median offset is less than $0.03''$ in both the right ascension and declination
directions for both fields. There is also no sign of any distortion or rotation over the fields. 
We astrometrically aligned all the other images to the $K_s$-band images. 
Thus, we can directly measure NIR and optical magnitudes at the radio source positions.

For each X-ray source, we identified the nearest $K_s$-band counterpart,
if there was one, or, otherwise, we applied an average astrometric offset (determined from every
X-ray source that had a $K_s$-band counterpart) to determine coordinates.
It is these coordinates that we take to be the X-ray source positions subsequently
(i.e., we use them when obtaining spectra or when measuring fluxes in other wavebands).

\subsection{Optical Spectroscopy}
\label{z}

In addition to our existing spectra for the X-ray sources in the CDF-N (Trouille et al.\ 2008)
and the publicly available spectra for both fields 
(e.g., Cohen et al.\ 2000; Cowie et al.\ 2004b; Szokoly et al. 2004; Wirth et al.\ 2004; Le F{\`e}vre et al.\ 2005; 
Reddy et al.\ 2006; Popesso et al.\ 2009; Treister et al. 2009; Balestra et al.\ 2010;
Silverman et al.\ 2010; Cooper et al.\ 2011, 2012),
we obtained new spectra for both fields
using the DEIMOS (Faber et al.\ 2003)
and MOSFIRE (McLean et al.\ 2012) spectrographs on the Keck 10~m telescopes
(e.g., Cowie et al.\ 2012).
H. Suh et al.\ (2015, in preparation) also obtained new NIR spectra on the CDF-S with the FMOS
(Kimura et al.\ 2010) spectrograph on the Subaru 8.2~m telescope.
There are many publicly available optical and NIR redshifts for the COSMOS field
(e.g., Lilly et al.\ 2007; Trump et al.\ 2007; Civano et al.\ 2012; Silverman et al.\ 2015).

For the {\em Chandra\/} sources in any of the X-ray bands in the CDF-S (740 from Xue et al.\ 2011)
and the CDF-N (503 from Alexander et al.\ 2003), we visually inspected all of the available
optical/NIR spectra to determine whether there is a secure redshift and to look for AGN 
signatures. 
We classified the sources with AGN signatures as broad-line AGNs (BLAGNs), where some lines in the spectrum
have full-width half maximum (FWHM) widths greater than 2000 km s$^{-1}$ (note that one object is a broad 
absorption line quasar or
BALQSO), and Seyfert type~2 sources (Sy2), where high excitation narrow lines are present 
(usually CIV$\lambda1549$, CIII]$\lambda1909$, or [NeV]$\lambda3426$; see, e.g., Szokoly et al.\ 2004). 
We refer to the sources without AGN signatures as ``other".
We note that the spectral classifications can sometimes be affected by the available wavelength coverage 
for the sources.

\subsection{X-ray Luminosities}
\label{xraylum}

Our primary results will come from the combined CDF-N and CDF-S observed-frame $4-8$~keV 
sample, but we will also consider the CDF-N sample alone when looking at the radio properties,
since we only have an ultradeep radio image for the CDF-N (Section~\ref{radio}).
For sources without spectroscopic redshifts, 
we use the photometric redshifts from Rafferty et al.\ (2011), where available. 

We calculate the rest-frame $8-16$~keV luminosities, $L_X$, from the observed-frame $4-8$~keV
fluxes using
\begin{equation}
L_X=4\pi d_L^2 f_{\rm 4-8~keV}~\Biggl({{1+z}\over 2}\Biggr)^{-0.2} \,.
\end{equation}
This is exact at $z=1$ but assumes an intrinsic photon index of $\Gamma=1.8$ to compute 
the $K-$corrections at other redshifts. We computed the rest-frame $1-4$~keV luminosities 
in the same way using the $0.5-2$~keV fluxes. In subsequent figures, we use these rest-frame
labels.
We adopt $L_X=10^{42}$~erg~s$^{-1}$ as a conservative threshold for AGN activity.
A source with $L_X>5\times10^{43}$~erg~s$^{-1}$ would be described as an X-ray quasar 
following the usual $L_{\rm 2-8~keV}>10^{44}$~erg~s$^{-1}$ definition based on the rest-frame 
$2-8$~keV flux (e.g., Barger et al.\ 2005).

In Figure~\ref{xray_props}, we plot optical spectral class versus $L_X$ for the combined
CDF-N and CDF-S observed-frame $4-8$~keV sample with spectroscopic or photometric redshifts.
The data show the now well-known effect that the fraction
of BLAGNs drops rapidly with decreasing X-ray luminosity 
(e.g., Cowie et al.\ 2003; Steffen et al.\ 2003; Ueda et al.\ 2003; Barger et al.\ 2005; 
Hasinger et al.\ 2005; La Franca et al.\ 2005).
At $L_X>5\times10^{43}$~erg~s~$^{-1}$, roughly one third  of the sources
are BLAGNs, while in the range $L_X=10^{43}-5\times10^{43}$~erg~s$^{-1}$,
this has dropped to just 10$\%$.

\begin{inlinefigure}
\begin{center}
\includegraphics[width=3.75in]{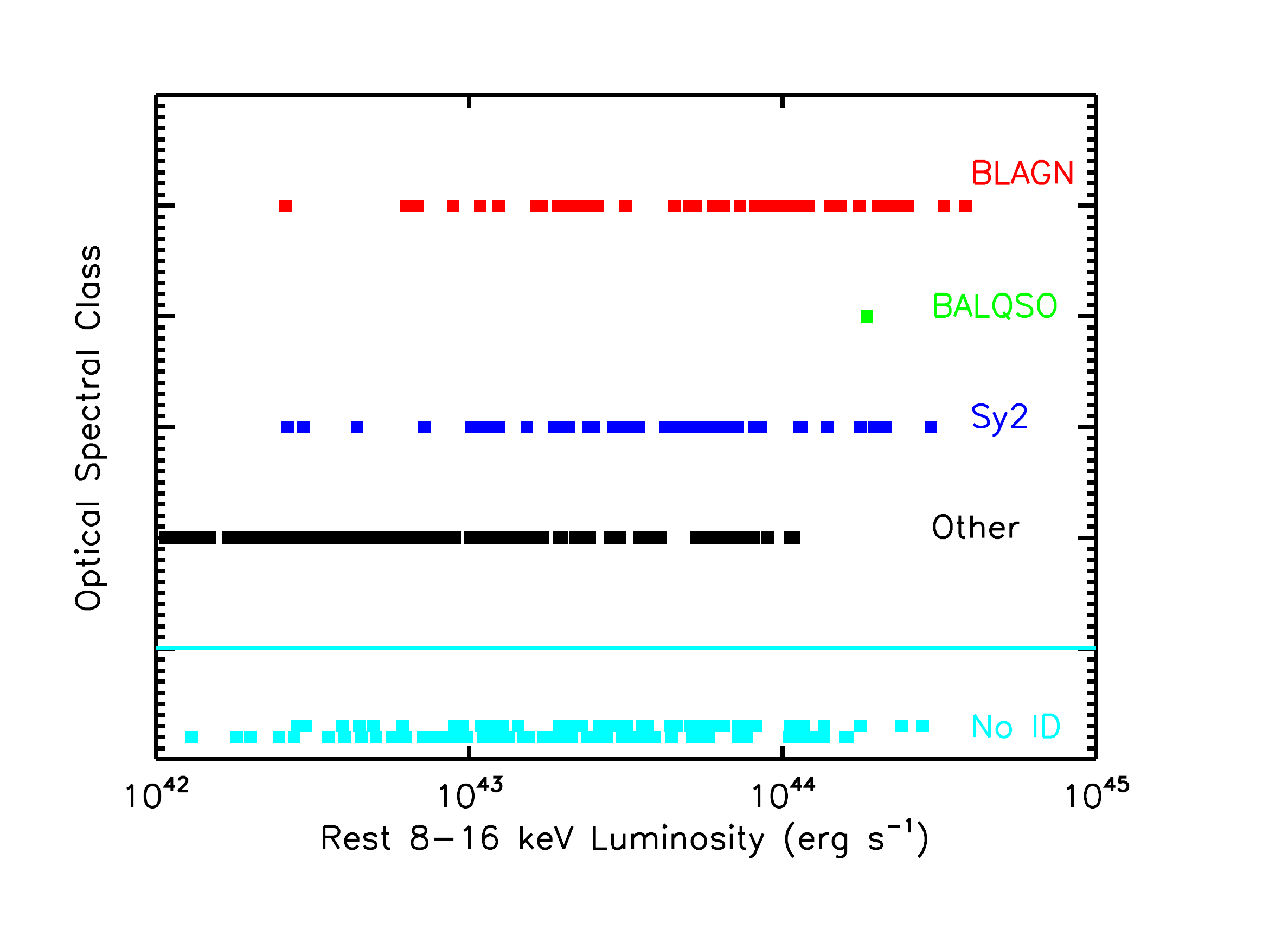}
\vskip -0.5cm
\caption{
Optical spectral class vs. $L_X$ for the observed-frame $4-8$~keV sources in the 
combined CDF-N and CDF-S fields with spectroscopic (red --- BLAGN; 
green --- BALQSO;  blue --- Sy2; black --- other) or photometric (cyan --- No ID;
plotted over two lines and below a horizontal demarcation for clarity) redshifts.
\label{xray_props}
}
\end{center}
\end{inlinefigure}

\subsection{Restricted X-ray Sample}
\label{xraysamp}

We further restrict the X-ray sample to only those X-ray sources that lie within 
a $6'$ off-axis angle
in each field, since that is where the X-ray observations are the most sensitive.  
(In each field, the X-ray flux limit rises to twice the central value at about $5\farcm3$.)
Fortunately, these areas roughly match the most sensitive areas observed 
with SCUBA-2 (Section~\ref{submm}) and are covered by deep millimeter 
observations from the ground, FIR 
observations from {\em Herschel\/}, and MIR observations from {\em Spitzer\/} MIPS
(Rieke et al.\ 2004) (Section~\ref{other}), and from
optical and NIR observations from {\em HST, Spitzer,\/} and the ground
(Sections~\ref{opt}). In total, these
areas contain 214 $4-8$~keV selected sources (98 from the CDF-N, and 116 from the CDF-S).
{\em Hereafter, we will refer to this as our combined CDF-N and CDF-S X-ray sample.}

In Figure~\ref{xray_types}(a), we plot redshift versus observed-frame $4-8$~keV flux
for the sample.
Two-thirds of the sources (140 out of the 214) have robust spectroscopic redshifts, while
63 of the remaining sources have photometric redshifts from Rafferty et al.\ (2011).
Eleven sources are too faint even for photometric redshifts. 

In Figure~\ref{xray_types}(b), we test the spectroscopic classifications against the X-ray 
spectral properties by plotting optical spectral class versus rest-frame flux ratio
 $(1-4~{\rm keV})/(8-16~{\rm keV})$ (i.e., observed-frame flux ratio $(0.5-2~{\rm keV})/(4-8~{\rm keV})$) 
for the 122 sources with spectroscopic (73) or photometric (49) redshifts $z>1$, which will be
the redshift range of most interest in this paper due to the sensitivities of the submillimeter
data. We can see from the figure that the BLAGNs
are all soft (the red solid line shows the mean value of the ratio for the BLAGNs, and the red dashed
line shows half that value to indicate the range covered)
and substantially disjoint from the Sy2 and BALQSO sources.

\vskip -1.0cm
\begin{inlinefigure}
\begin{center}
\centerline{\includegraphics[width=3.75in]{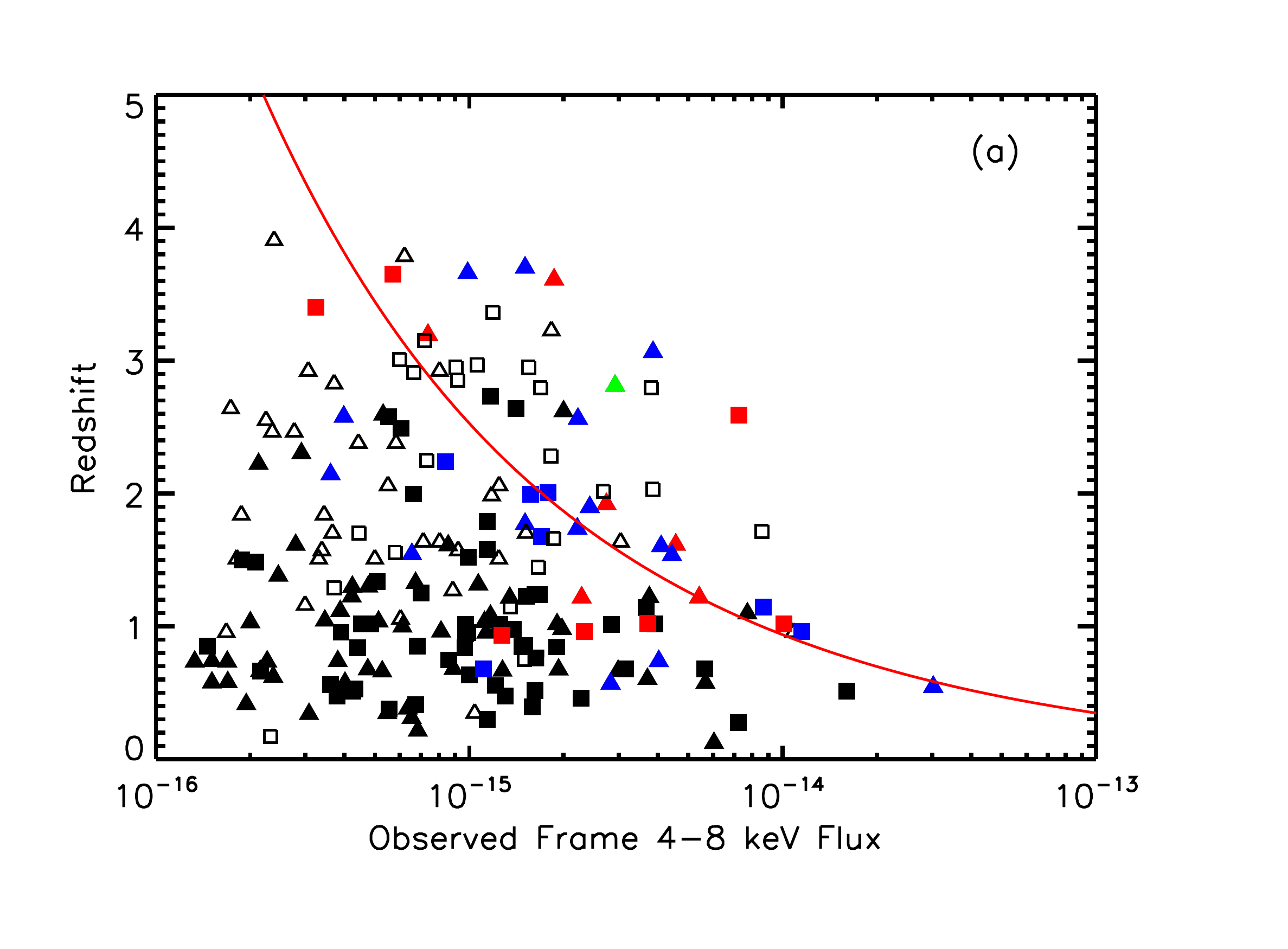}}
\vskip -1.0cm
\centerline{\includegraphics[width=3.75in]{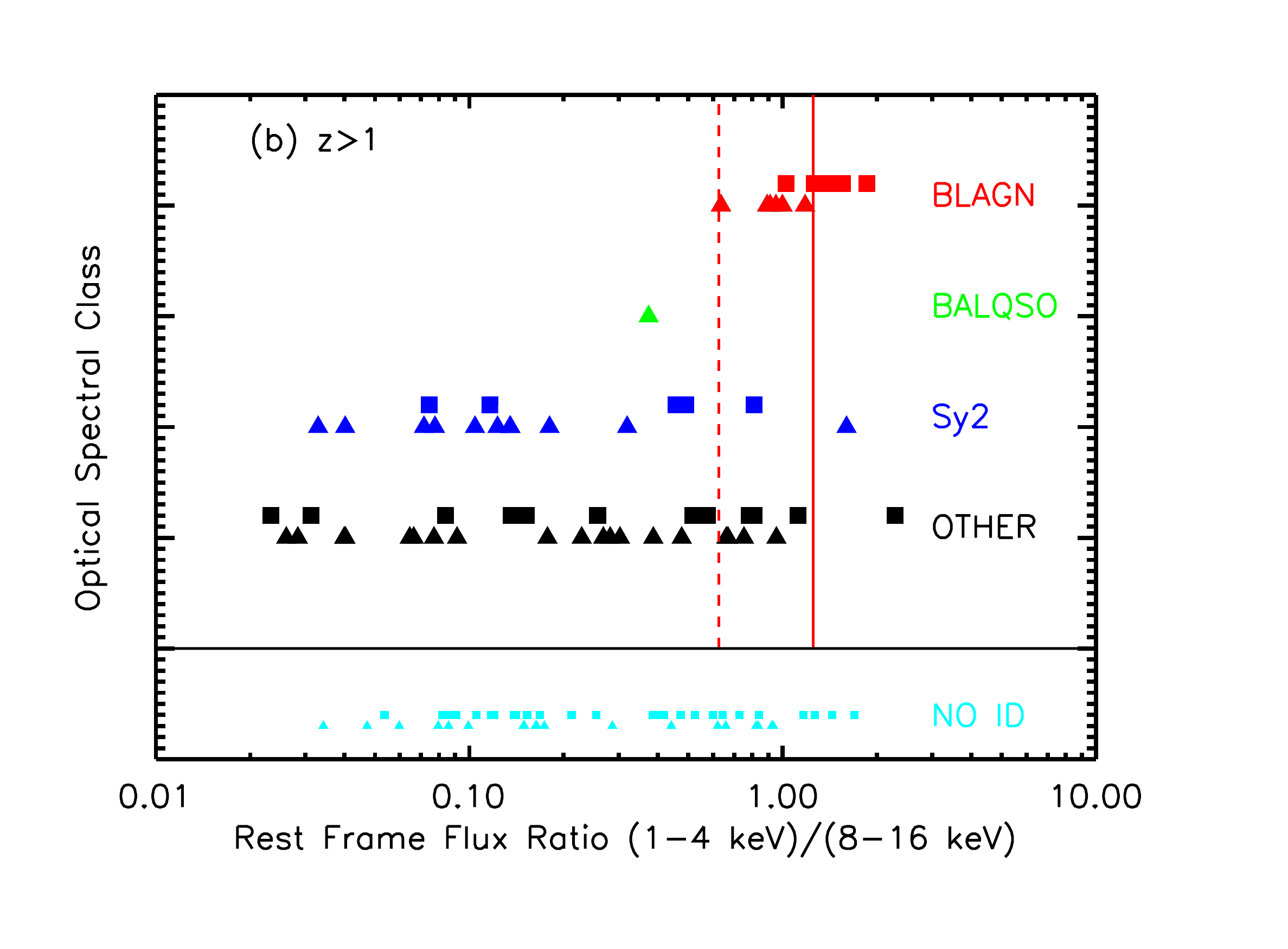}}
\vskip -0.5cm
\caption{
(a) Redshift vs. observed-frame $4-8$~keV flux for the X-ray sources in the $6'$ radius 
regions of the CDF-N (squares) and CDF-S (triangles) with spectroscopic 
(red --- BLAGN; green --- BALQSO;  blue --- Sy2; black --- other) 
or photometric (open --- No ID) redshifts.
(b) Optical spectral class vs. rest-frame flux ratio
$(1-4~{\rm keV})/(8-16~{\rm keV})$
for the same sources but restricted to $z>1$.
In (a), the red curve corresponds to $L_X=5\times 10^{43}$~erg~s$^{-1}$. 
In (b), the red solid line shows the mean value for  
the BLAGNs in the sample, and the red dashed line shows half that value.
\label{xray_types}
}
\end{center}
\end{inlinefigure}

\subsection{Radio Imaging}
\label{radio}

In order to study the location of the X-ray sources on the FIR-radio and MIR-radio correlations,
we use the extremely deep 1.4~GHz image of the CDF-N field obtained by 
F. Owen (2015, in preparation).
The image covers a $40'$ diameter region with an effective resolution of
$1\farcs8$. The absolute radio positions are known to $0\farcs1-0\farcs2$ rms.
The highest sensitivity region is about $9'$ in radius. Thus, in the core $6'$ radius region
that we chose for the X-ray sample (Section~\ref{xraysamp}),
the radio map is relatively uniform with an rms of 2.3~$\mu$Jy.
There are 447 distinct $>5\sigma$ radio sources in the core region, excluding sources that appear
to be parts of other sources.
{\em Hereafter, we will refer to this as our radio sample.}

Matching counterparts from the $>5\sigma$ radio 
catalog to the X-ray sources is not critically dependent on the choice 
of match radius (Alexander et al.\ 2003), so we followed Barger et al.\ (2007) and chose a 
search radius of $1\farcs5$. Of the 417 $2-8$~keV sources within a $9'$ off-axis radius
in the Alexander et al.\ sample, 195 have counterparts in the $5\sigma$ radio sample, 
while of the 260 $2-8$~keV sources within the 
$6'$ off-axis angle (where the X-ray data are deeper and hence the X-ray sources fainter), 
125 have radio counterparts. For our $4-8$~keV sample, 55 of the 98 sources within the 
chosen $6'$ radius have radio counterparts.
In the following, we will use the measured radio fluxes from F. Owen (2015, in
preparation) for the matched sources, and we will
measure radio fluxes at the X-ray source positions for the remaining unmatched X-ray sources.

\subsection{Millimeter, FIR, and MIR Imaging}
\label{other}

In order to construct the FIR SEDs, we use our SCUBA-2 images 
(Section~\ref{submm}) and the publicly available images listed in Table~\ref{tabFIR}. 
Note that the AzTEC 1.1~mm (Perera et al.\ 2008) and MAMBO 1.2~mm (Greve et al.\ 2008)
data in the GOODS-N were combined into a deeper map at an effective
wavelength of 1.16~mm by Penner et al.\ (2011), and the 100~$\mu$m and 160~$\mu$m
PACS Evolutionary Probe (PEP) (P.I. D. Lutz; described in Lutz et al.\ 2011) 
and GOODS-H data were combined into a deeper image by Magnelli et al.\ (2013),
so it is the combined image references that we give in the table.

We performed point spread function (PSF) weighted smoothing on all of the images, since we do not 
expect the high-redshift sources to be resolved.
We then measured the fluxes of the sources in the smoothed images
at their X-ray and radio source positions. A small number (9) of sources in the X-ray sample 
and a small number (10) of sources in the radio sample
were excluded at this stage, because the sources lie too close to very bright FIR sources.
For each source in each image, we measured the confusion noise by measuring the fluxes at random
positions that had comparable sensitivities using an identical procedure.

\subsection{SCUBA-2 Imaging}
\label{submm}

We obtained 51.3~hours of SCUBA-2 observations on the CDF-N and 49.6~hours 
on the CDF-S during observing 
runs in 2012, 2013, and 2014 (Chen et al.\ 2013b; Barger et al.\ 2014; 
L. Cowie et al.\ 2015, in preparation).
Details of the observational procedures and data reduction of these data using the 
Dynamic Iterative
Map-Maker (DIMM) in the SMURF package from the STARLINK software developed
by the Joint Astronomy Centre (Chapin et al.\ 2013) may be found in Chen et al.\ (2013b). 

The noise maps are obtained by computing the variance of the data that lands in
each pixel. Chen et al.\ (2013a) tested the robustness of the noise maps created by
DIMM by checking whether the standard deviation of the signal-to-noise ratio (S/N) 
maps are close to 1.
They confirmed that the noise maps are accurate with an underestimation less than
5\%. They postulated that the cause of the small underestimation is the correlated noise
from the large-scale structure caused by the atmospheric noise, which ultimately gets 
subtracted out in the post-processing (see their Section 2.1).

\vskip 0.5cm
\begin{inlinefigure}
\hskip 2.5cm
\centerline{\includegraphics[width=3.2in,angle=0]{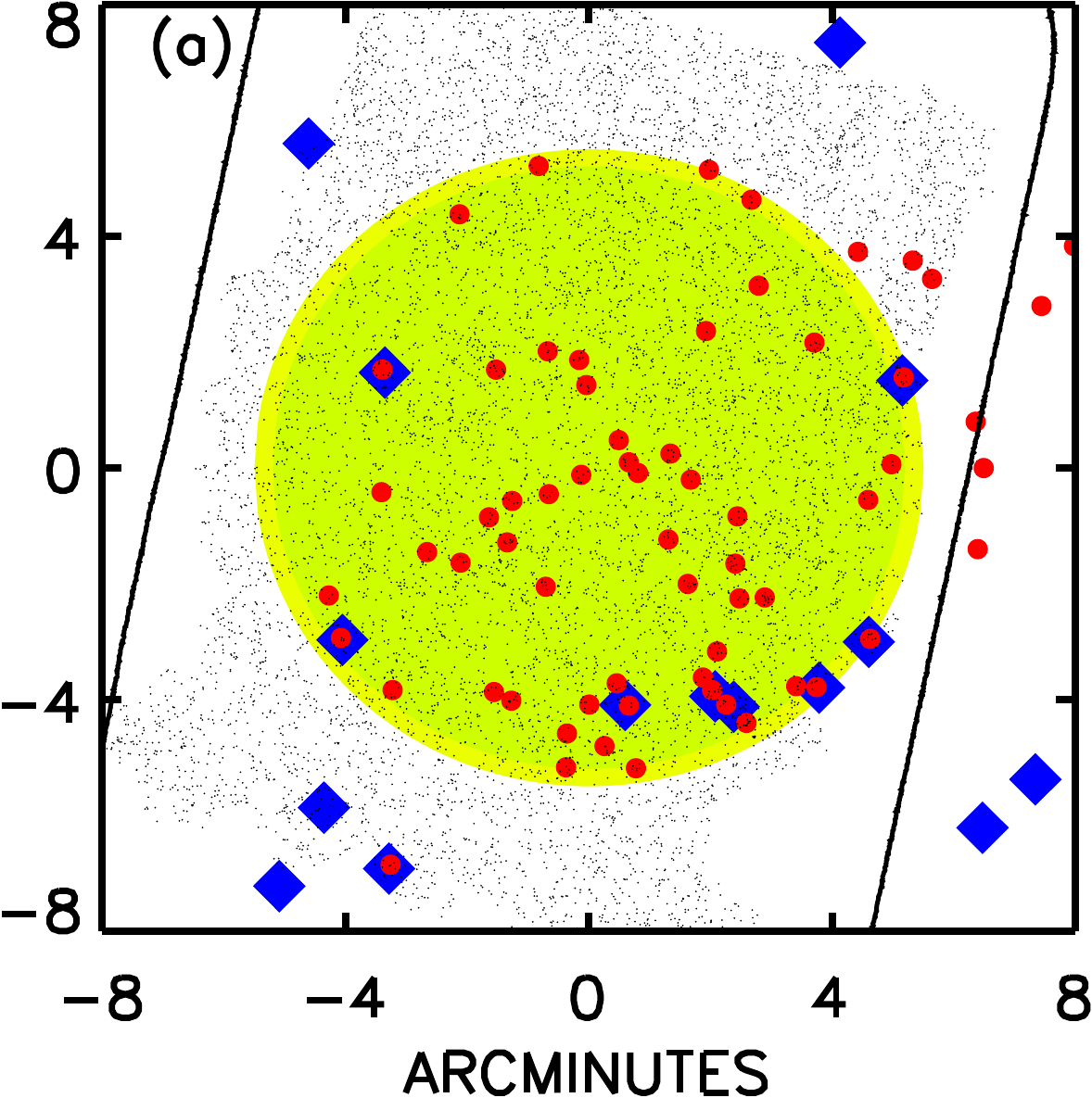}}
\vskip 0.15cm
\hskip 2.5cm
\centerline{\includegraphics[width=3.2in,angle=0]{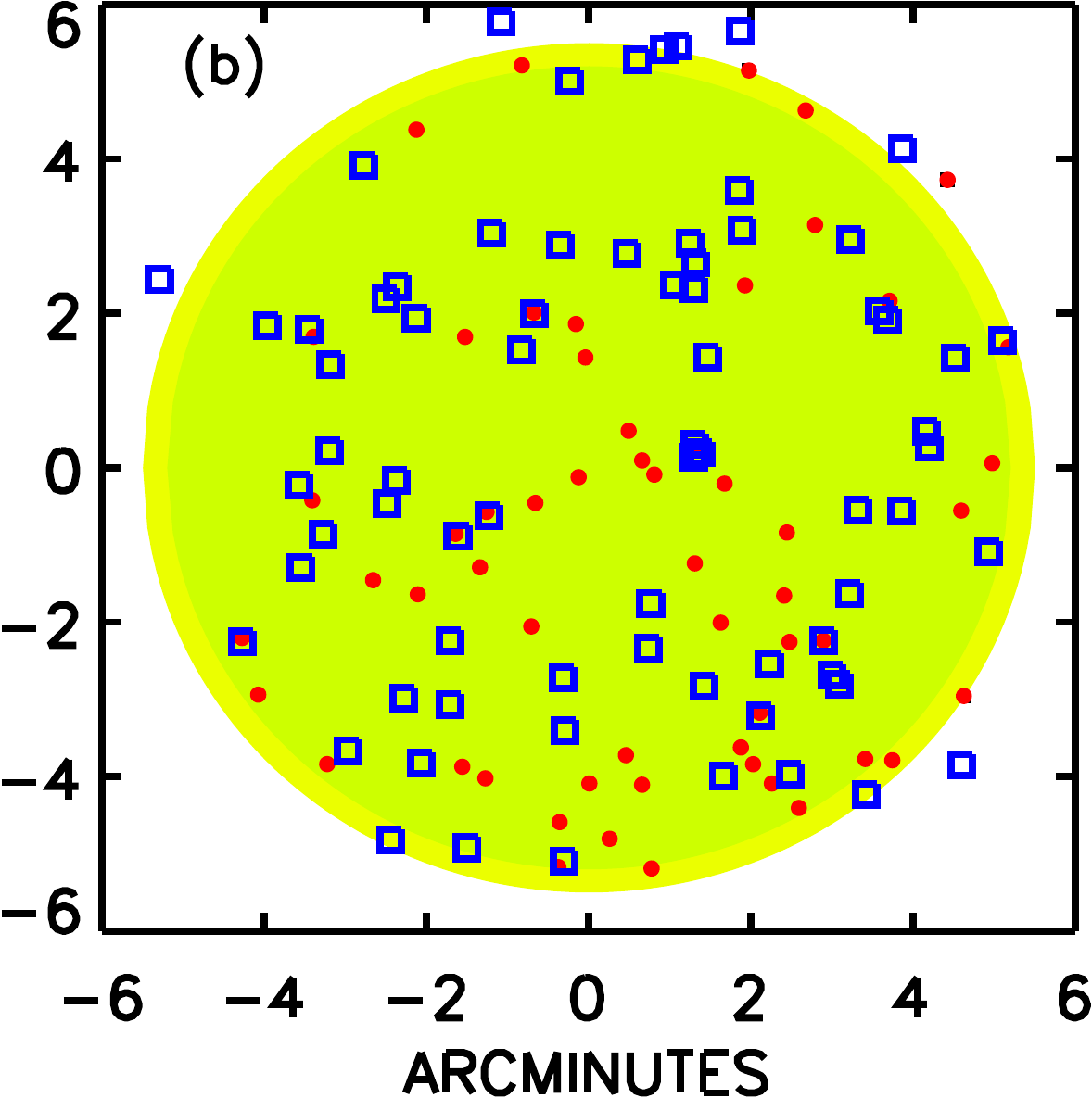}}
\caption{
(a) Deep areas (less than twice the central noise) in both the
X-ray (light green shading) and SCUBA-2 (yellow shading) observations of the
CDF-S. We also compare our directly detected SCUBA-2 $4\sigma$ sources
(red circles) with previous observations of the field from the LABOCA survey (blue diamonds). 
The black contour shows where the PEP+GOODS-H 100~$\mu$m exposure time
exceeds 5\% of the maximum exposure time, while the gray shading shows the 
CANDELS coverage.
(b) Image from (a) expanded to show the positions 
of the  $z>1$ observed-frame $4-8$~keV sample (blue squares) plotted on top of
the SCUBA-2 sources (red circles). 
\label{image}
}
\end{inlinefigure}

\begin{deluxetable}{llll}
\renewcommand\baselinestretch{1.0}
\tablewidth{0pt}
\tablecaption{Millimeter, FIR, and MIR Imaging}
\scriptsize
\tablehead{$\lambda$ ($\mu$m)          &  Instr./Tel.   & Field & Reference \\
(1) & (2) & (3) & (4) }
\startdata
1160 & AzTEC/JCMT & CDF-N & Penner et al.\ (2011) \cr
&+MAMBO/IRAM & \cr
500, 350, 250 & SPIRE/{\em Herschel\/} & CDF-N & Oliver et al.\ (2012) \cr
500, 350, 250 & SPIRE/{\em Herschel\/} & CDF-S & Elbaz et al.\ (2011) \cr
160, 100	& PACS/{\em Herschel\/} 	& 	CDF-N & Magnelli et al.\ (2013) \cr
& & CDF-S & \cr
70                  		& PACS/{\em Herschel\/} 	&	CDF-S & Lutz et al.\ (2011) \cr
24 & MIPS/{\em Spitzer\/} & CDF-N & P.I. M. Dickinson \cr
& & CDF-S &
\enddata
\label{tabFIR}
\end{deluxetable}

The data were primarily obtained with the CV DAISY scanning
modes in band~1 ($\tau_{225~{\rm GHz}}$ opacity $<0.05$) or band~2
($\tau_{225~{\rm GHz}}$ opacity $\sim0.05-0.08$) weather conditions.
(Detailed information about the SCUBA-2 scan patterns can be found in
Holland et al.\ 2013.)
CV DAISY is optimal for going deep on small areas,
such as the central deep regions of the two {\em Chandra\/} fields.
The central rms 850~$\mu$m sensitivity is 0.37~mJy  in both the CDF-N
and the CDF-S, but it increases with off-axis angle where the coverage 
becomes sparser.

The SCUBA-2 observations on the COSMOS field
by Casey et al.\ (2013) had an exposure time of 38.0~hours 
and were obtained using the PONG-900 scan pattern in band~1 weather conditions.
They covered 281.7~arcmin$^2$ with 
a uniform rms 850~$\mu$m sensitivity of 0.87~mJy.

For each field, we formed a matched filter image by weighting
the SCUBA-2 image with the PSF. This provides an optimal
estimate of the flux at any position provided that, as expected,
the sources are small compared with the beam full width half
maximum (FWHM) of $14''$ at 850~$\mu$m. We used a wider filter to subtract
variable backgrounds so that the average measured
flux at random positions in the image equals zero. 
For a detailed description of the reduction and calibration of SCUBA-2 data,
we refer the reader to Chapin et al.\ (2013) and Dempsey et al.\ (2013).

For each image, we generated a catalog of sources lying above a 4$\sigma$ threshold. 
There are 68 directly detected SCUBA-2 sources in the CDF-N, most of whose properties
are described in Barger et al.\ (2014), 64 in the CDF-S 
(L. Cowie et al.\ 2015, in preparation), and 99 in COSMOS (Casey et al.\ 2013).

In Figure~\ref{image}(a), we compare the region of the
SCUBA-2 image of the CDF-S field where the rms sensitivity is less than twice the central 
noise limit (yellow shading; roughly a $5\farcm5$ radius area)
with the region of the {\em Chandra\/} image where the sensitivity is also less
than twice the central noise limit (light green shading; $5\farcm3$ off-axis angle). 
We can see that the sensitive region of the SCUBA-2 image provides a reasonable match to the 
sensitive region of the X-ray image, as well as to the sensitive region of the PEP+GOODS-H image
(solid contour shows the region with exposure times in excess of 5\% of the maximum 
exposure time) and to the {\em HST\/} CANDELS image  (gray shading).

In Figure~\ref{image}(a),
we also compare our $>4\sigma$ SCUBA-2 850~$\mu$m sample
(red circles) with the $>4\sigma$ LABOCA 860~$\mu$m sample from
Wei\ss\ et al.\ (2009, blue diamonds). The latter 
sample was the basis for the ALMA follow-up survey that was used in Wang et al.\ (2013) to study 
the X-ray fraction and X-ray properties of submillimeter galaxies.
However, our survey, which is much deeper, yields a large sample of directly
detected submillimeter galaxies in the deep regions of the 4~Ms {\em Chandra\/} exposure, 
including many near on-axis sources, while the LABOCA survey does not.
(We find 55 SCUBA-2 sources within the $5\farcm5$ radius region, while LABOCA finds only 8;
see Chen et al.\ 2013b for a discussion.)

We may use the SMA observations in the CDF-N field (e.g., Wang et al. 2011; Barger et al.\ 2012, 2014;
some new observations) to test the positional accuracy of the SCUBA-2 data. 
Thirty of the 32 SMA sources lie within $3\farcs6$ of their
SCUBA-2 counterpart. The rms offset is $1\farcs9$ between
the SMA and SCUBA-2 positions. In what follows, we use
a conservative radius of $4''$ in matching other samples
to the directly detected SCUBA-2 sources.

In Figure~\ref{image}(b), we show an expansion of Figure~\ref{image}(a) to compare
the observed-frame $4-8$~keV band $z>1$ sources (blue squares)
with the $>4\sigma$ SCUBA-2 sources (red circles) in the CDF-S.
Six of these X-ray sources lie within a $4''$ match radius from a 
SCUBA-2 source (none of which is also a LABOCA source). 

Ten of the 119 sources in the combined CDF-N and CDF-S sample that lie at $z>1$ have 
SCUBA-2 counterparts using the $4''$ match radius criterion. 
There is only one match in the COSMOS sample, given its much shallower $4-8$~keV data.

\section{Dusty Star Formation Signatures in the X-ray Sample}
\label{SF}

We begin our analysis with the submillimeter observations.
Over the redshift range $z=1-5$, the 850~$\mu$m observed wavelength corresponds to
rest-frame wavelengths greater than 142~$\mu$m.
It is generally thought that at these wavelengths, the light is produced
primarily by star formation with little contribution from the AGN 
(e.g., Fritz et al.\ 2006; Netzer et al.\ 2007; Hatziminaoglou et al.\ 2010; Mullaney et al.\ 2012). 
Thus, while the exact conversion from monochromatic flux to FIR luminosity, and hence to SFR, 
for a given source depends on its SED, we may directly determine whether sources in the X-ray
sample have star formation signatures from the submillimeter observations alone. 
Here we perform a statistical analysis on the distribution function of 850~$\mu$m fluxes
measured for the X-ray sources.
We will turn to the full FIR SEDs in the next two sections.

In measuring the 850~$\mu$m fluxes, we first removed all of the directly detected $>4\sigma$ 
SCUBA-2 sources from the matched filter smoothed images using a PSF based on the observed 
calibrators. This left residual images from which we measured the 850~$\mu$m fluxes
(whether positive or negative) and statistical errors at the X-ray source positions. The only
exceptions were if there existed an X-ray counterpart to a $>4\sigma$ SCUBA-2 source,
as described in Section~\ref{submm}.
In these cases, we assigned the SCUBA-2 source flux to the X-ray counterpart, since  
the flux could not be correctly measured from the residual image.
This approach minimizes the contamination of the fainter submillimeter 
sources by the brighter ones, at the expense of assuming the counterpart 
matches from Section~\ref{submm} are correct.
Our comparison with the CDF-N SMA observations showed that
most of the counterpart matches are indeed correct;
however, in a small number of cases where the SCUBA-2 source is
a blend of fainter sources, we may have misidentifications.

In Figure~\ref{lx_smm}(a), we show the measured 850~$\mu$m fluxes and $1\sigma$ 
statistical errors versus $L_X$ for the combined CDF-N and CDF-S X-ray sample with 
either spectroscopic or photometric redshifts $z>1$ (black squares).
We find that 13 of the X-ray sources ($\sim 10$\% of the sample) are detected 
above the $3\sigma$ level at 850~$\mu$m (red squares). 
(Note that we use a $4\sigma$ detection threshold for our direct SCUBA-2 detections,
but we can lower this to $3\sigma$ when using a pre-determined sample selected
at another wavelength, in this case X-rays.)
The bulk of the detected sources are intermediate $L_X$ sources.

We determined the contamination (the number of spurious $3\sigma$ 
detections produced by neighboring objects) and the confusion error by generating Monte Carlo 
randomized positions for each source in areas of the image that had comparable
sensitivities to those at the source's position.
The Monte Carlo results give an average false detection rate of 1 source
and a 95$\%$ confidence upper limit of 2 sources, so nearly all of
the $3\sigma$ detections in Figure~\ref{lx_smm}(a) are real.

\begin{inlinefigure}
\centerline{\includegraphics[width=3.70in]{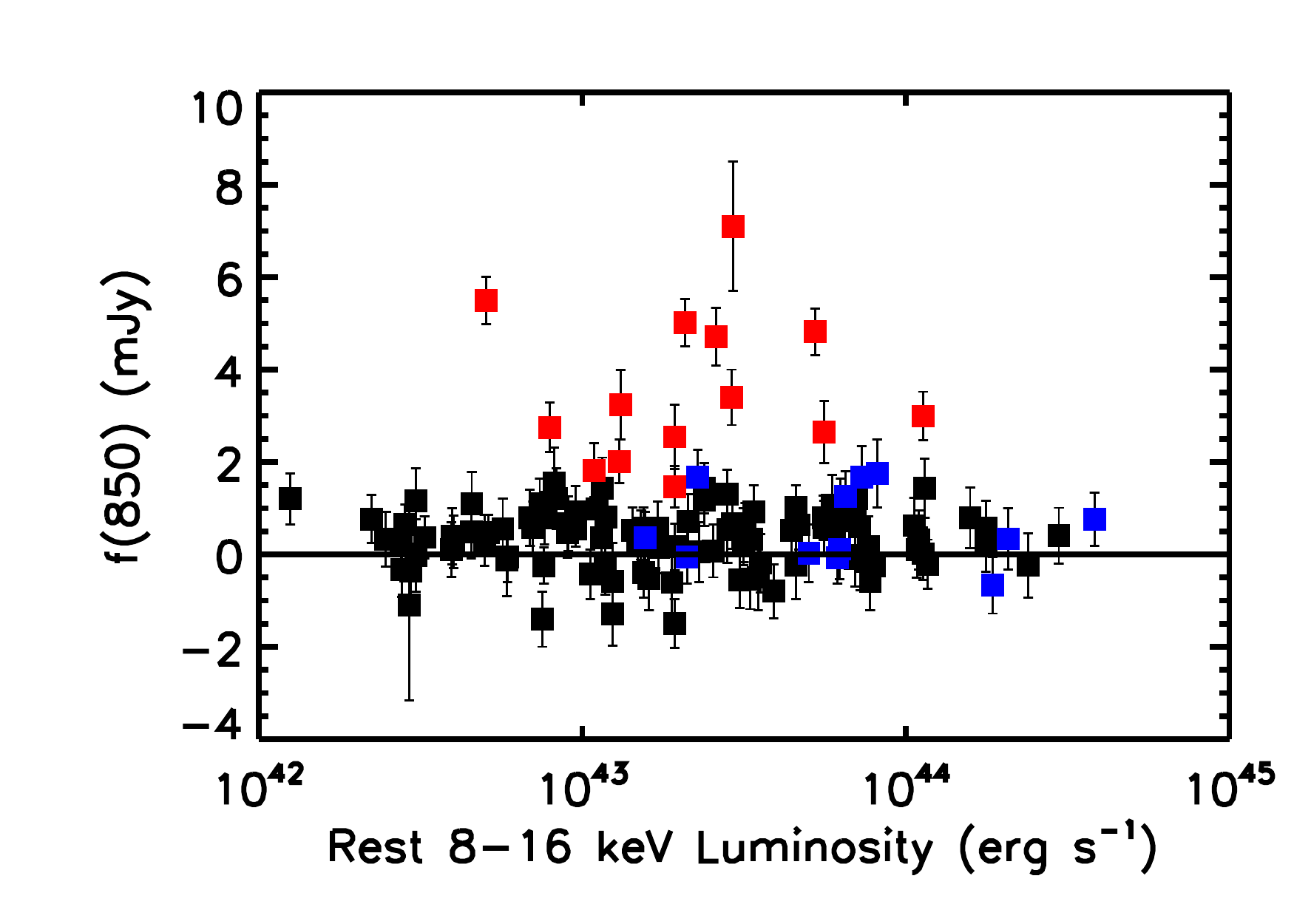}}
\centerline{\includegraphics[width=3.25in]{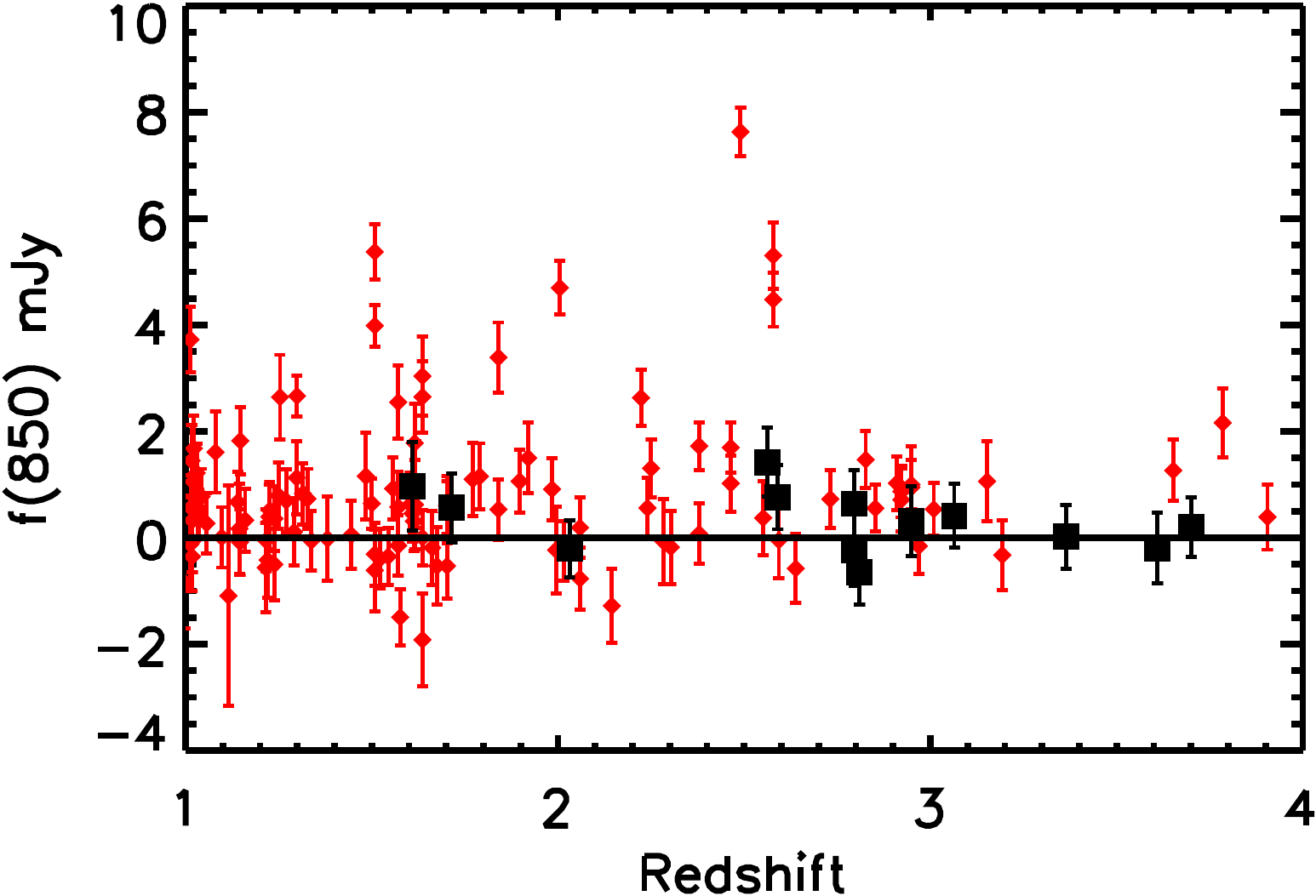}}
\caption{
(a) 850~$\mu$m flux vs. $L_X$ for the combined CDF-N and CDF-S X-ray sample with either 
spectroscopic or photometric redshifts $z>1$ (black squares).
The red squares denote the X-ray sources detected above the $3\sigma$ level at 850~$\mu$m.
This is a lower detection threshold than what we used for our direct SCUBA-2 detections
($4\sigma$), since here we have a pre-determined sample selected at another wavelength.
The blue squares denote sources that are BLAGNs. The error
bars are $1\sigma$ statistical errors. 
(b) 850~$\mu$m flux vs. redshift for the same sample. 
The red diamonds show sources with
$L_X\le10^{44}$~erg~s$^{-1}$, while the black squares show those with 
$L_X>10^{44}$~erg~s$^{-1}$. The error bars are $1\sigma$ statistical errors.
\label{lx_smm}
}
\end{inlinefigure}

The $L_X>10^{44}$~erg~s$^{-1}$ sources have
an error-weighted mean 850~$\mu$m flux of 0.48$\pm0.11$~mJy, 
while for the $L_X=10^{43}-10^{44}$~erg~s$^{-1}$
sources, it is 0.92$\pm0.06$~mJy. 
If we assume an Arp~220 SED at $z>1$, then this means that the 
$L_X=10^{43}-10^{44}$~erg~s$^{-1}$
sources lie in hosts with average SFRs of $\sim185~M_\odot$~yr$^{-1}$,
while the $L_X>10^{44}$~erg~s$^{-1}$ sources lie in hosts with average
SFRs that are roughly two times lower. 

We note that none of the 18 BLAGNs (blue squares) are detected directly. 
To test the robustness of  this result on a larger sample, we expanded the 
present CDF-N and CDF-S X-ray
sample to include regions where the submillimeter errors are higher. 
This resulted in submillimeter flux measurements for 57 BLAGNs, 
only 2 of which have detections above the $3\sigma$ level.
The error-weighted mean 850~$\mu$m flux for the 57 BLAGNs is $0.59\pm0.14$~mJy,
while the error-weighted mean flux of the Sy2s having $L_X>10^{43}$~erg~s$^{-1}$ is 
$1.04\pm0.12$~mJy.

This result differs from submillimeter observations of luminous optical quasar samples. 
For example, the combined samples of Priddey et al.\ (2003) and Omont et al. (2003)
contain 83 quasars with $M_B<-27.5$ (Orellana et al.\ 2011). 
Of these, 18 were detected at $>3\sigma$ and had
either $850~\mu$m fluxes (Priddey et al.) in the range $7-17$~mJy or 
1.2~mm fluxes (Omont et al.) in the range $3.2-10.7$~mJy.
The two $>3\sigma$ detected sources in our sample are less luminous, with both having 
850~$\mu$m fluxes of only 2.6~mJy. Thus, we do not have any sources in
our sample that would have been detected by these previous surveys. 
The fraction of submillimeter detected BLAGNs in our expanded sample is 0.04 (0.01, 0.08), 
which can be compared with 0.22 for the bright optical quasar sample. 
Thus, there is a significant difference between the BLAGNs in this sample and the
more luminous objects in the bright optical quasar sample.

In Figure~\ref{lx_smm}(b), we show the measured 850~$\mu$m fluxes and $1\sigma$ statistical
errors versus redshift for the combined CDF-N and CDF-S X-ray sample with either spectroscopic 
or photometric redshifts $z>1$.
In this figure, we have color-coded the sources by X-ray luminosity 
(red diamonds denote $L_X\le10^{44}$~erg~s$^{-1}$,
and black squares denote $L_X>10^{44}$~erg~s$^{-1}$).
We see no clear evidence for a dependence on redshift;
however, the $L_X\le10^{44}$~erg~s$^{-1}$ sources
have a much wider spread of 850~$\mu$m flux at a given redshift
than the $L_X>10^{44}$~erg~s$^{-1}$ sources.
A Kolmogorov-Smirnov test shows only a 0.036 probability that the two samples are
drawn from the same distribution.

\begin{deluxetable}{cccc}
\renewcommand\baselinestretch{1.0}
\tablewidth{0pt}
\tablecaption{Mean 850~$\mu$m Fluxes}
\scriptsize
\tablehead{$\log L_X$ Interval & Mean Flux & Error & Number \\ 
(ergs~s$^{-1}$) & (mJy) & (mJy) & (mJy) \\
(1) & (2) & (3) & (4) }
\startdata
42.0-42.5 & 0.42    & 0.18   &   11    \cr
42.5-43.0 & 0.84    & 0.11   &   41    \cr
43.0-43.5 & 1.09    & 0.085   &   82     \cr
43.5-44.0 & 0.70    & 0.09   &   81     \cr
44.0-44.5 & 0.49    & 0.13   &   57     \cr
44.5-45.0 & 0.37    & 0.44   &   4
\enddata
\label{tabMEANS}
\end{deluxetable}

In Figure~\ref{lx_smm_means}(a), we show the statistical error-weighted mean 850~$\mu$m fluxes
in six half-dex intervals of $L_X$ for all three fields, both separately (CDF-N, red squares;
CDF-S, green diamonds; COSMOS, blue triangles) and all together (black large squares).
We note the number of sources in each interval at the bottom of the figure.

In Table~\ref{tabMEANS}, we provide the numerical values for the combined sample, including
the $L_X$ interval (Column~1), the error-weighted mean 850~$\mu$m flux (Column~2) and
$1\sigma$ statistical error (Column~3) in that interval, and the number of sources in that interval
(Column~4).

\begin{inlinefigure}
\begin{center}
\includegraphics[width=3.75in]{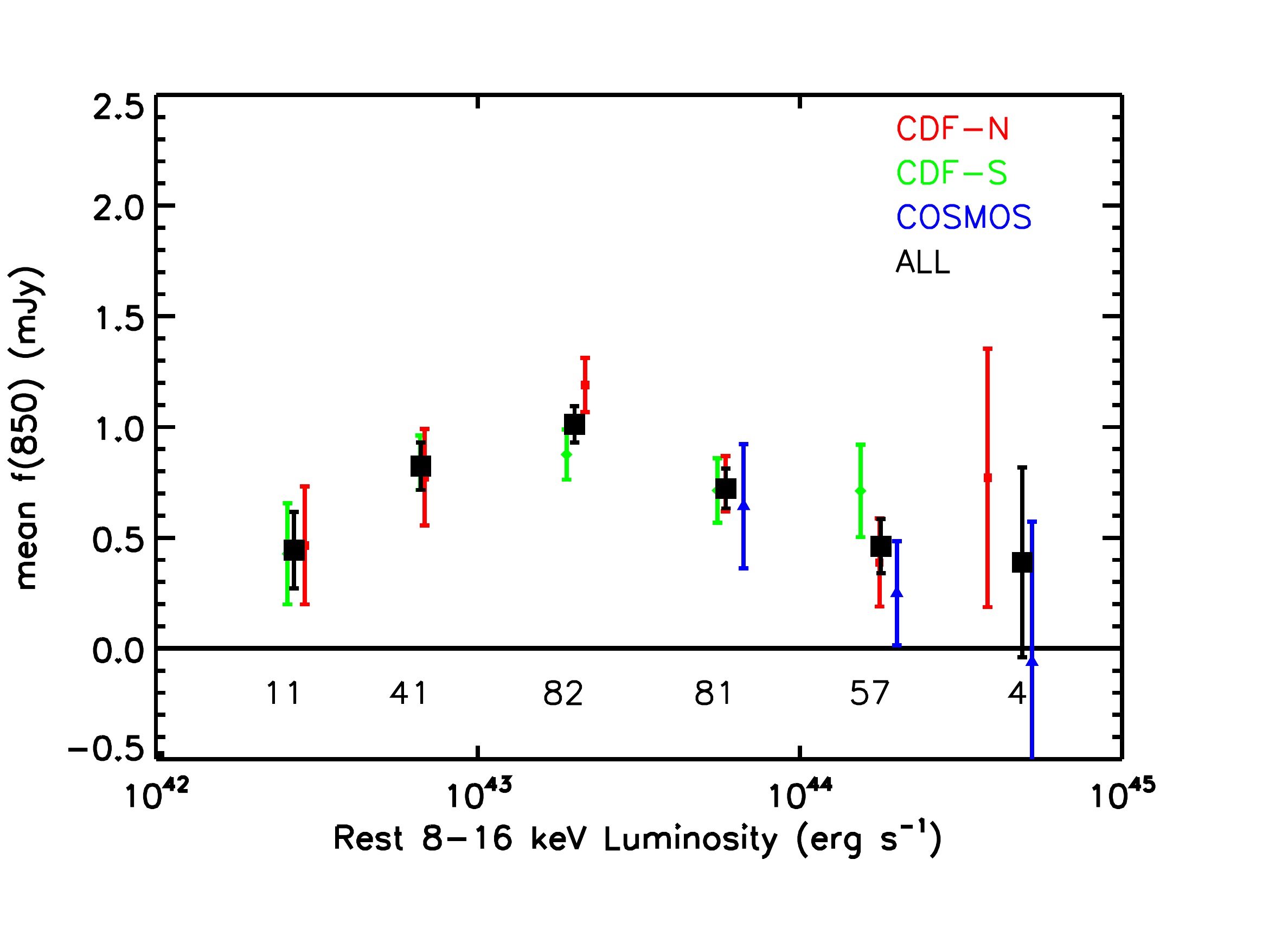}
\vskip -0.75cm
\includegraphics[width=3.75in]{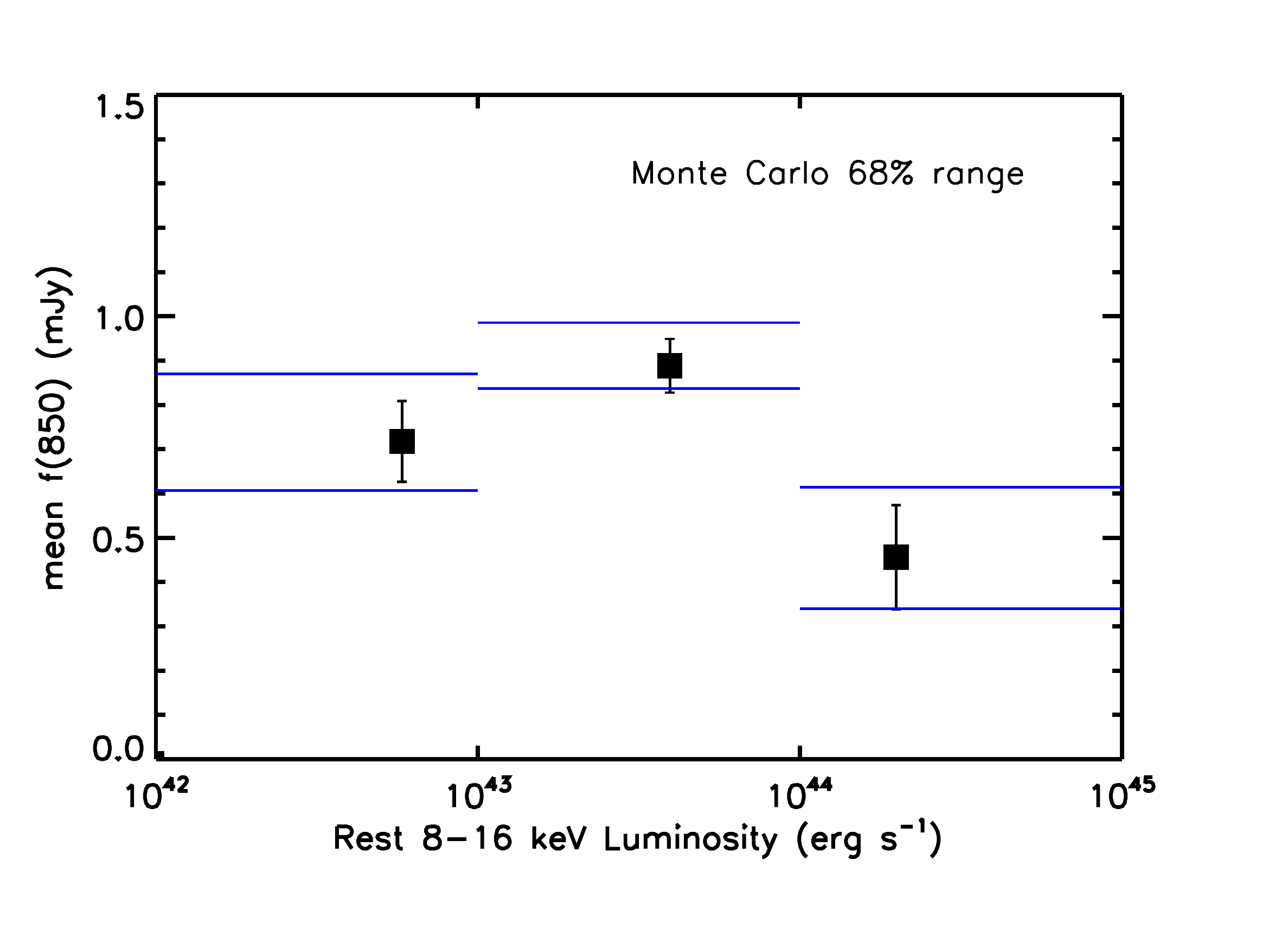}
\vskip -0.5cm
\caption{
(a) Statistical error-weighted mean 850~$\mu$m fluxes in six half-dex intervals of $L_X$
for the individual fields (red squares---CDF-N; green diamonds---CDF-S; 
blue triangles---COSMOS) and all together (black large squares). The points
are plotted at the mean luminosities in each interval with statistical error bars.
The number of sources in each interval is given at the bottom.
(b) Statistical error-weighted mean 850~$\mu$m fluxes in three one-dex intervals of $L_X$
for all the fields together. The points
are plotted at the mean $L_X$ in each interval with statistical error bars.
The 68\% confidence ranges determined from the 
Monte Carlo simulations are also shown (blue horizontal lines).
\label{lx_smm_means}
}
\end{center}
\end{inlinefigure}

We can see that there is a rise in the error-weighted mean 850~$\mu$m fluxes with $L_X$,
reaching a peak at the intermediate $L_X$ values of
$10^{43}-10^{43.5}$~erg~s$^{-1}$, followed by a decline to higher $L_X$ values.
Thus, provided that the FIR light is indeed predominately 
produced by star formation in the host galaxy rather than by accretion onto the central AGN, 
we can infer that for sources at low and 
intermediate X-ray luminosities, the host galaxy SFRs rise with AGN luminosity,
but for the high X-ray  luminosities, the SFRs decline.

The question of interest here is whether the lower mean 850~$\mu$m fluxes for the high X-ray 
luminosity sources relative to the intermediate X-ray luminosity sources is statistically significant. 
In Figure~\ref{lx_smm_means}(b), we show the statistical error-weighted mean 850~$\mu$m fluxes
in three one-dex intervals of $L_X$ for all the fields together.
Here we compare the statistical error bars
with the 68\% confidence ranges determined from the Monte Carlo simulations (blue horizontal lines). 
The error bars from the Monte Carlo simulations are only
very slightly larger than the statistical errors that dominate the uncertainties.
The Monte Carlo simulations give only a 0.013 probability that the 850~$\mu$m
fluxes drawn from the $L_X>10^{44}$~erg~s$^{-1}$ population are as large
as those drawn from the $L_X=10^{43}-10^{44}$~erg~s$^{-1}$ population.

 \section{Constructing SEDs for the X-ray Sample}
 \label{SED}

From the wealth of existing data on the CDF-N and CDF-S fields, we are now able to construct 
SEDs for the X-ray sample to investigate how the observed dependence of submillimeter flux 
on X-ray luminosity from Section~\ref{SF} is reflected in the FIR shapes.
In Section~4.1, we will focus on providing a visualization of the data
using stacked SEDs as a function of X-ray luminosity and redshift, while in
Section~\ref{indivfits}, we will fit each X-ray (and radio) source's SED individually.

To measure the 850~$\mu$m fluxes here, we use a different approach than we used in 
Section~\ref{SF}:  We simply measure the fluxes from the matched filter smoothed SCUBA-2 
images without first removing all of the directly detected $>4\sigma$ SCUBA-2 sources.
We measure the fluxes in the other FIR bands in the same way. Although this
results in higher levels of contamination and confusion than one gets from using maps
that have been cleaned using priors, such as the 850~$\mu$m or 24~$\mu$m
source positions and fluxes, it avoids assumptions about the relation of
the priors to the data in each bandpass and allows for a much simpler
statistical analysis. To deal with the contamination/confusion, we use measurements made at
Monte Carlo randomized positions to determine the background and errors for each
X-ray source.

 \subsection{Average SEDs}
 \label{avgSED}

To construct the average SEDs, we first interpolate the measured $L_\nu \nu$
values for each source onto a common rest-frame
wavelength vector. We then form a simple average in each bin.
In Figure~\ref{stack_sed_lx}, we show the average $L_\nu \nu$ SEDs (colored curves) 
in four $L_X$ intervals (we combine the two highest $L_X$ intervals and the
two lowest $L_X$ intervals from Table~\ref{tabMEANS})
for sources with redshifts (a) $z=2.0-4.5$, (b) $z=0.8-2.0$, and (c) $z=0.4-0.8$.
(Note that we do not have any $L_X\le10^{43}$~erg~s$^{-1}$ sources in the highest redshift 
interval due to our X-ray flux limits, and there are no $L_X>10^{44}$~erg~s$^{-1}$
sources in the lowest redshift interval.)
We show with shading the 68$\%$ confidence range computed using the bootstrap method
for the redshift-luminosity intervals with $\ge10$ sources.  Otherwise, we show only the mean
(e.g., black curve in (b); green and purples curves in (c)).

We  denote the mean $L_X$ for each redshift-luminosity interval with a horizontal line of the
same color. (These mean values can be read off the right-hand axis.)
This makes it possible to compare the bolometric luminosities at each wavelength with $L_X$ 
across the wavelength range. For example, in (b), the ratio of the bolometric luminosity at $1~\mu$m
to $L_X$ for the cyan curve is high due to the stellar contribution. 
In contrast, the green curve is nearly completely AGN dominated at that wavelength.

\begin{inlinefigure}
\centerline{\includegraphics[width=3.25in]{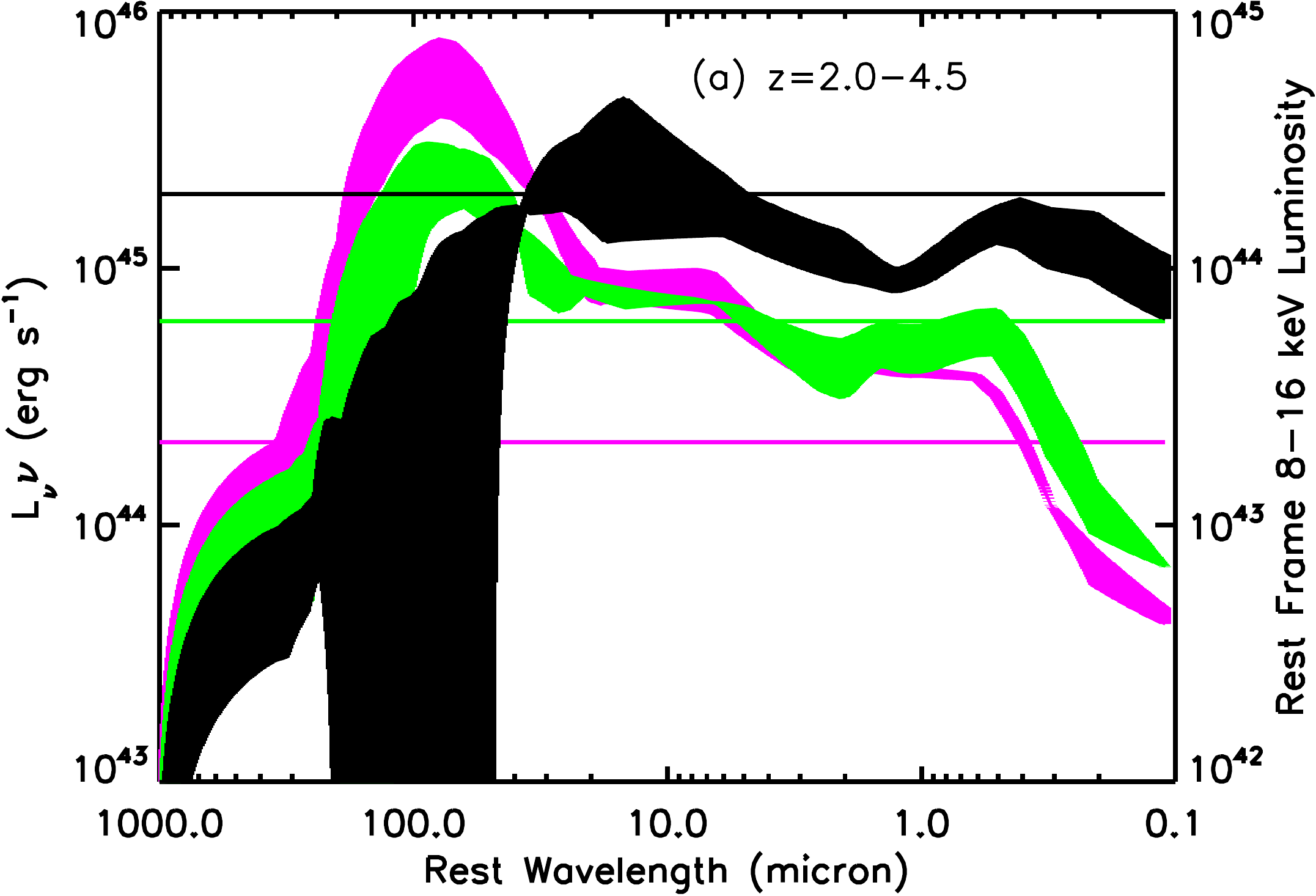}}
\vskip 0.5cm
\centerline{\includegraphics[width=3.25in]{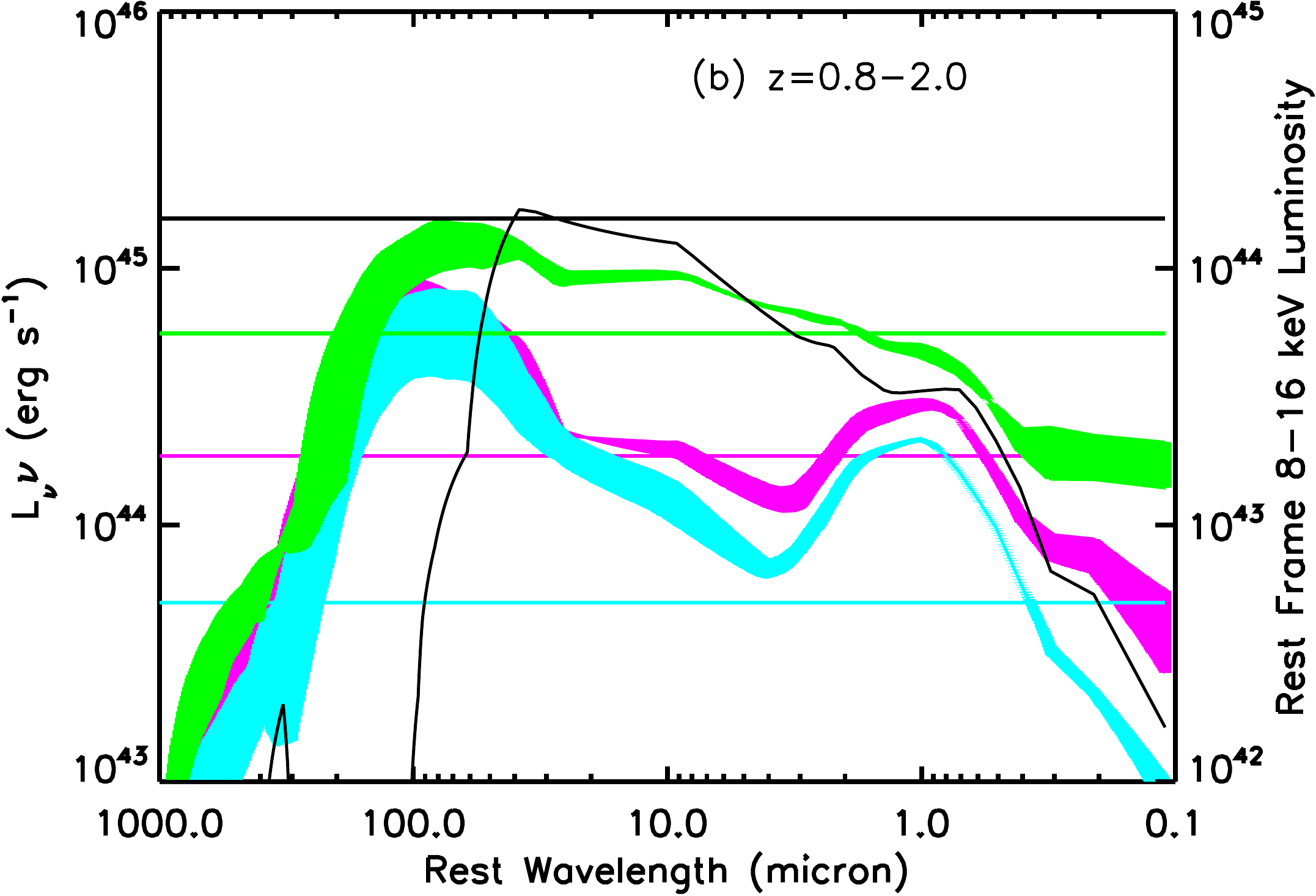}}
\vskip -1.0cm
\hskip -0.5cm
\centerline{\includegraphics[width=4.25in]{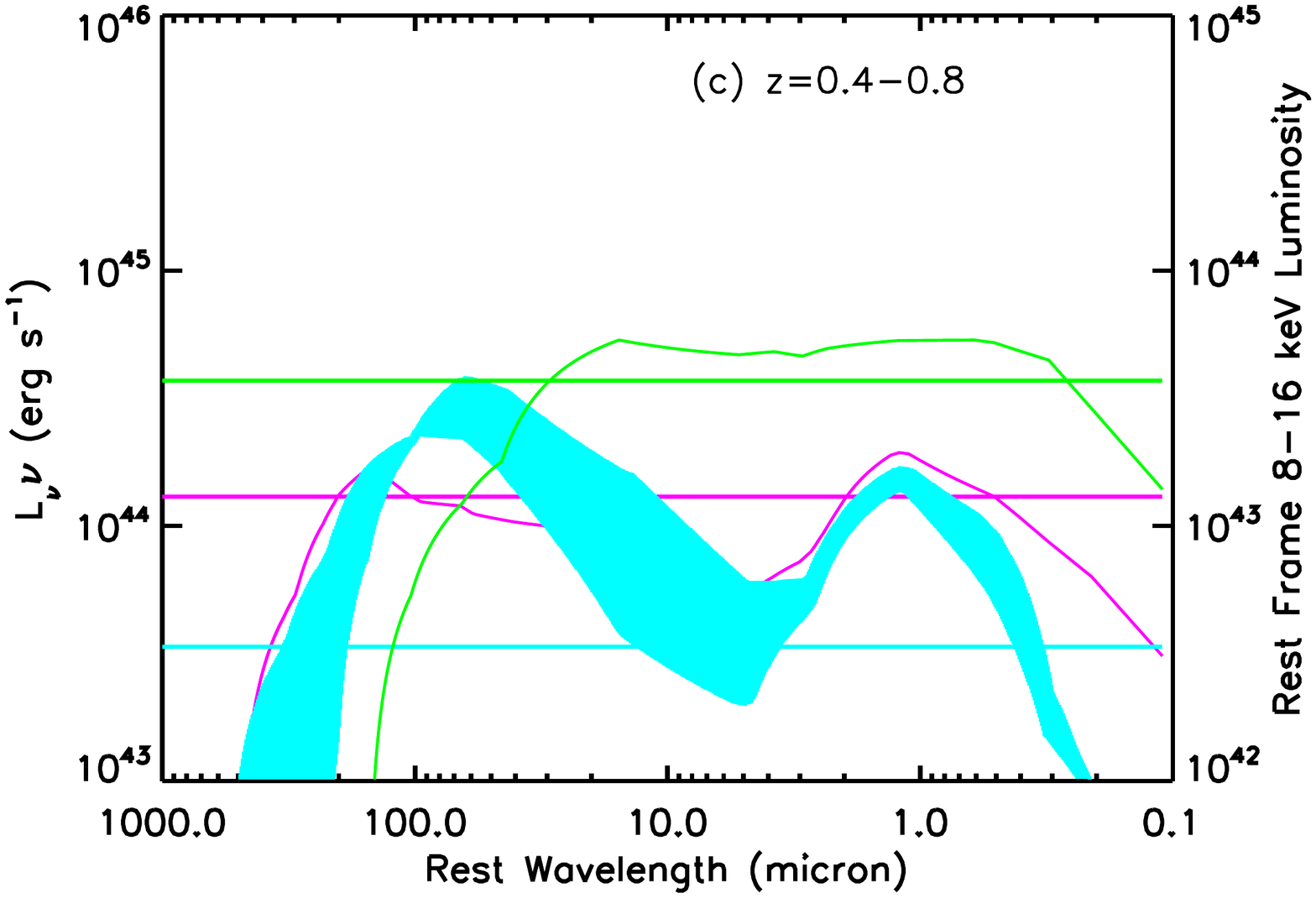}}
\vskip -1.0cm
\caption{
Average $L_\nu \nu$ SEDs (colored curves) in four rest-frame $8-16$~keV luminosity intervals 
(black --- $L_{X}>10^{44}$~erg~s$^{-1}$; green --- $L_{X}=10^{43.5}-10^{44}$~erg~s$^{-1}$;
purple --- $L_{X}=10^{43}-10^{43.5}$~erg~s$^{-1}$; cyan --- $L_{X}=10^{42}-10^{43}$~erg~s$^{-1}$)
for the X-ray sources with redshifts
(a) $z=2.0-4.5$, (b) $z=0.8-2.0$, and (c) $z=0.4-0.8$.
The numbers of sources in each redshift interval are  (a) black --- 10;
green --- 16; purple --- 18; (b) black --- 3; green --- 15; purple --- 30; cyan --- 53;
(c) green --- 1; purple --- 3; cyan --- 20.
The bands show the 68$\%$ confidence intervals calculated with the bootstrap
method for the redshift-luminosity intervals with $\geq 10$ sources. 
Otherwise, only the mean is plotted.
The mean $L_X$ values (right-hand axis) of the sources in each X-ray luminosity
interval are also shown (horizontal lines).
\label{stack_sed_lx}
}
\end{inlinefigure}

\begin{inlinefigure}
\centerline{\includegraphics[width=3.5in]{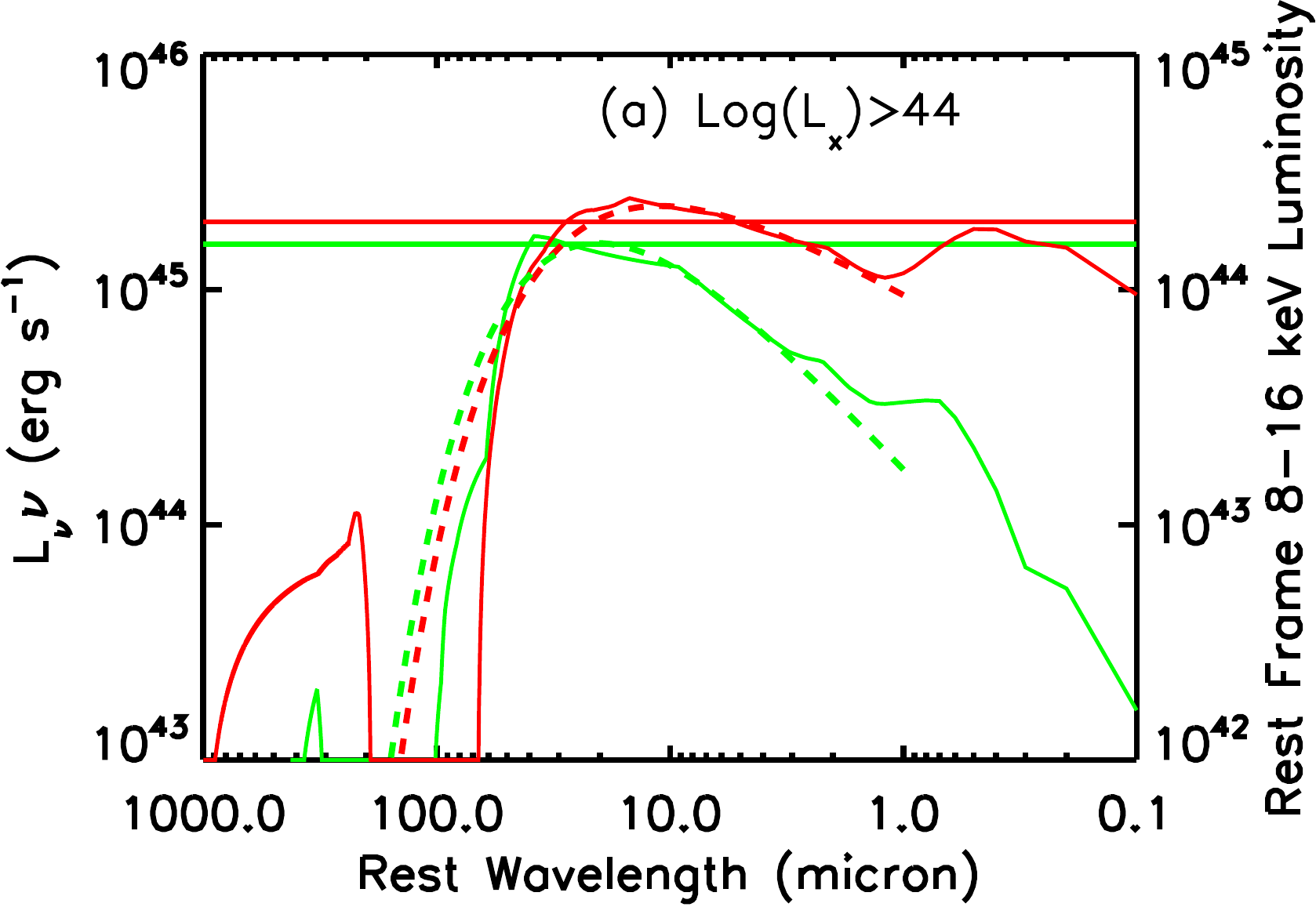}}
\centerline{\includegraphics[width=3.75in]{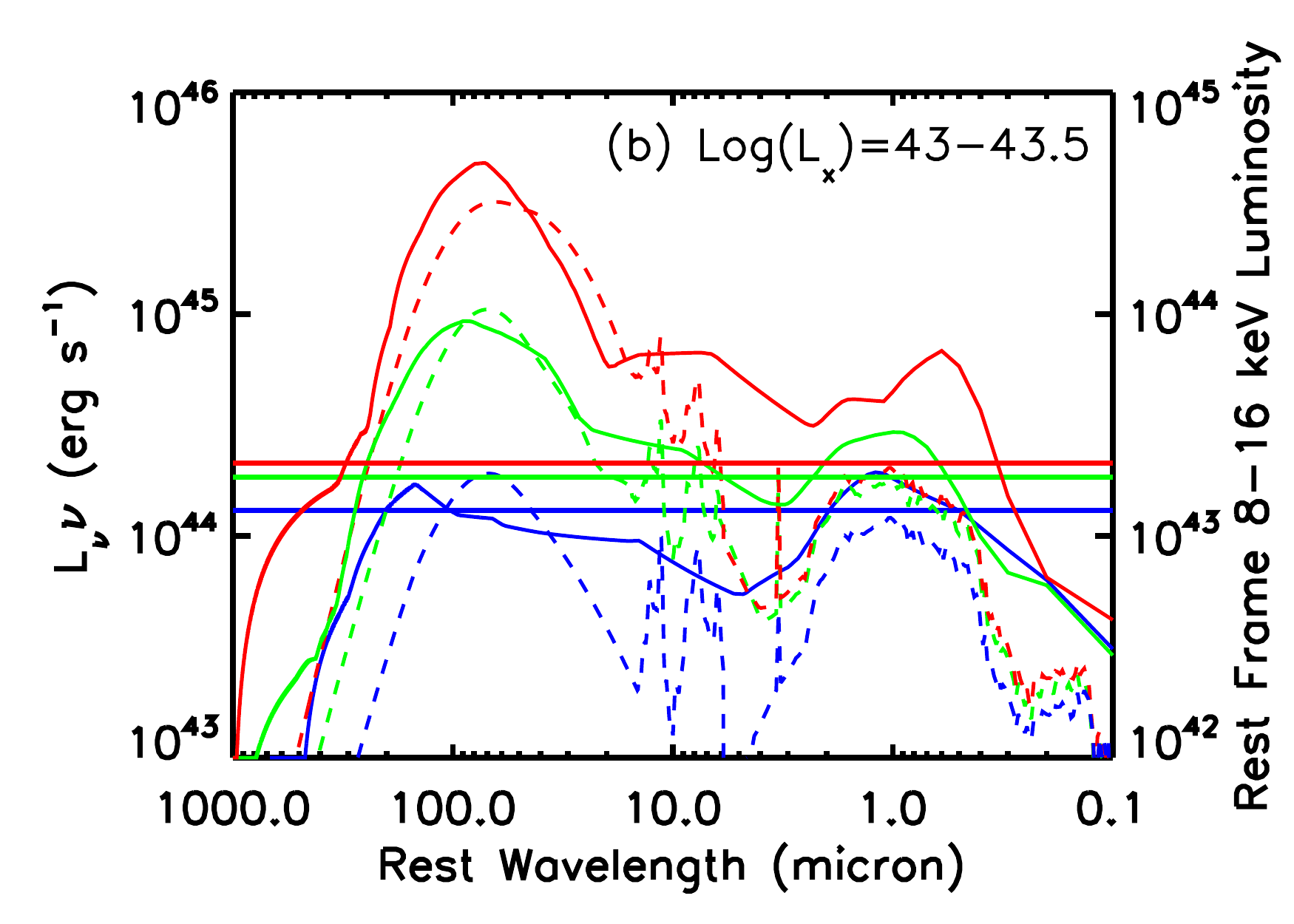}}
\caption{
Average $L_\nu \nu$ SEDs (solid curves) in the rest-frame $8-16$~keV luminosity intervals
(a) $L_{X}>10^{44}$~erg~s$^{-1}$ and
(b) $L_{X}=10^{43}-10^{43.5}$~erg~s$^{-1}$ for the redshift intervals
$z=2.0-4.5$ (red),  $z=0.8-2.0$ (green), and  $z=0.4-0.8$ (blue).
(Note that there are no sources at $z=0.4-0.8$ with $L_X>10^{44}$~erg~s$^{-1}$.)
The horizontal lines show the mean $L_X$ values of the sources in each  redshift 
interval (right-hand axis). In (a), the dashed curves show power law fits 
of the form $w^{\alpha} \times \exp(-w/w_0)$.
In (b), the dashed curves show Chary \& Elbaz (2001) templates
(see text for details).
\label{chary_sed}
}
\end{inlinefigure}

In Figure~\ref{chary_sed}, we show the average SEDs (solid curves)
for each of the three redshift intervals in
Figure~\ref{stack_sed_lx}, this time plotted according to X-ray luminosity interval (we show only 
two of the four):
(a) $L_X>10^{44}$~erg~s$^{-1}$ (these were the black curves in Figure~\ref{stack_sed_lx}) and 
(b) $L_X=10^{43}-10^{43.5}$~erg~s$^{-1}$ (these were the purple curves in Figure~\ref{stack_sed_lx}).
We again show the mean $L_X$ for each redshift-luminosity interval with a horizontal line of the same color.

In Figure~\ref{chary_sed}(a), the highest X-ray luminosity interval, we see that the 
FIR luminosities have vanished
for both redshift intervals (recall that the $z=0.4-0.8$ interval did not have any $L_X>10^{44}$~erg~s$^{-1}$ 
sources and hence is not shown). 
This observed long wavelength cut-off is another indication that the star formation needed to 
produce a substantial FIR luminosity is not taking place in the host galaxies 
of the most X-ray luminous sources. 

We show power law fits to the average SEDs with a long wavelength exponential 
cut-off of the form $w^{\alpha} \times \exp(-w/w_0)$ (dashed curves). 
Here $w$ is the wavelength and $w_0$ is 
the exponential scale length. We find values of
$\alpha=0.6$ and $w_0=19~\mu$m  for $z=2.0-4.5$ and $\alpha=1.1$ and $w_0=19~\mu$m  
for $z=0.8-2.0$.
We note that the actual cut-off  appears to be sharper at longer wavelengths than the 
exponential fall off gives.

In Figure~\ref{chary_sed}(b), we compare the average SEDs
with model templates from the Chary \& Elbaz (2001) SED library (dashed curves).
The Chary \& Elbaz templates are parameterized by their FIR
luminosities, $L_{\rm FIR}$, which are quoted for the
wavelength range $40-500~\mu$m and
follow the convention defined by Sanders \& Mirabel (1996). 
This means there is a unique SED shape for every FIR luminosity. 
We have chosen the templates that visually most closely match the FIR shapes and 
normalizations of our average SEDs.

For the lowest redshift interval (blue),
the Chary \& Elbaz template fit gives $L_{\rm FIR} \sim 8\times10^{10}~L_\odot$; 
this $L_{\rm FIR}$ roughly corresponds to the definition of a luminous 
infrared galaxy  or LIRG.
For the intermediate redshift interval (green), the template fit 
corresponds to $L_{\rm FIR} = 4\times10^{11}~L_\odot$, and for the
highest redshift interval (red), the template fit corresponds to $L_{\rm FIR} = 1.5\times10^{12}~L_\odot$,
an ultraluminous galaxy or ULIRG.
Thus, the FIR luminosities of the host galaxies are rising with increasing redshift in this
intermediate X-ray luminosity interval.

\subsection{Individual SED Fits}
\label{indivfits}
 
Motivated by the average SEDs of Section~\ref{avgSED}, as well as by theoretical considerations
(e.g., Fritz et al.\ 2006; Netzer et al.\ 2007), we next fit the individual source
SEDs at wavelengths longer than an observed-frame wavelength of $5.6~\mu$m 
(i.e., we include the {\em Spitzer\/} MIPS 24~$\mu$m data point and the two longest wavelength 
{\em Spitzer\/} IRAC data points in the fit) with a combined FIR 
gray body and MIR power law; we truncate the latter at the longer wavelengths using an exponential 
of the form $\exp^{-\lambda/(25~\mu{\rm m})}$. We used the Levinson-Marquardt based IDL fitting 
procedure of Markwardt (2009). The advantage of this type of fitting over that suggested by Casey (2012) 
is that it is simple to implement analytically. The fit contains 5 parameters: the slope ($\beta$), the 
temperature and normalization of the gray body, and the normalization and index of the power law.

As a check on the reliability of our fits, we performed a single gray body fit to only the {\em Herschel\/}
and submillimeter/millimeter data (constraining the gray body temperature to lie between 20 and 60~K)
without simultaneously fitting the MIR data with the truncated power law.
While the individual values for the gray body luminosity based on this fit can differ from the gray body
luminosities that we obtained from the combined gray body and truncated power law fit, we confirm 
that none of our subsequent results would change significantly if we were to use these values instead.

\begin{inlinefigure}
\centerline{\includegraphics[width=4.25in]{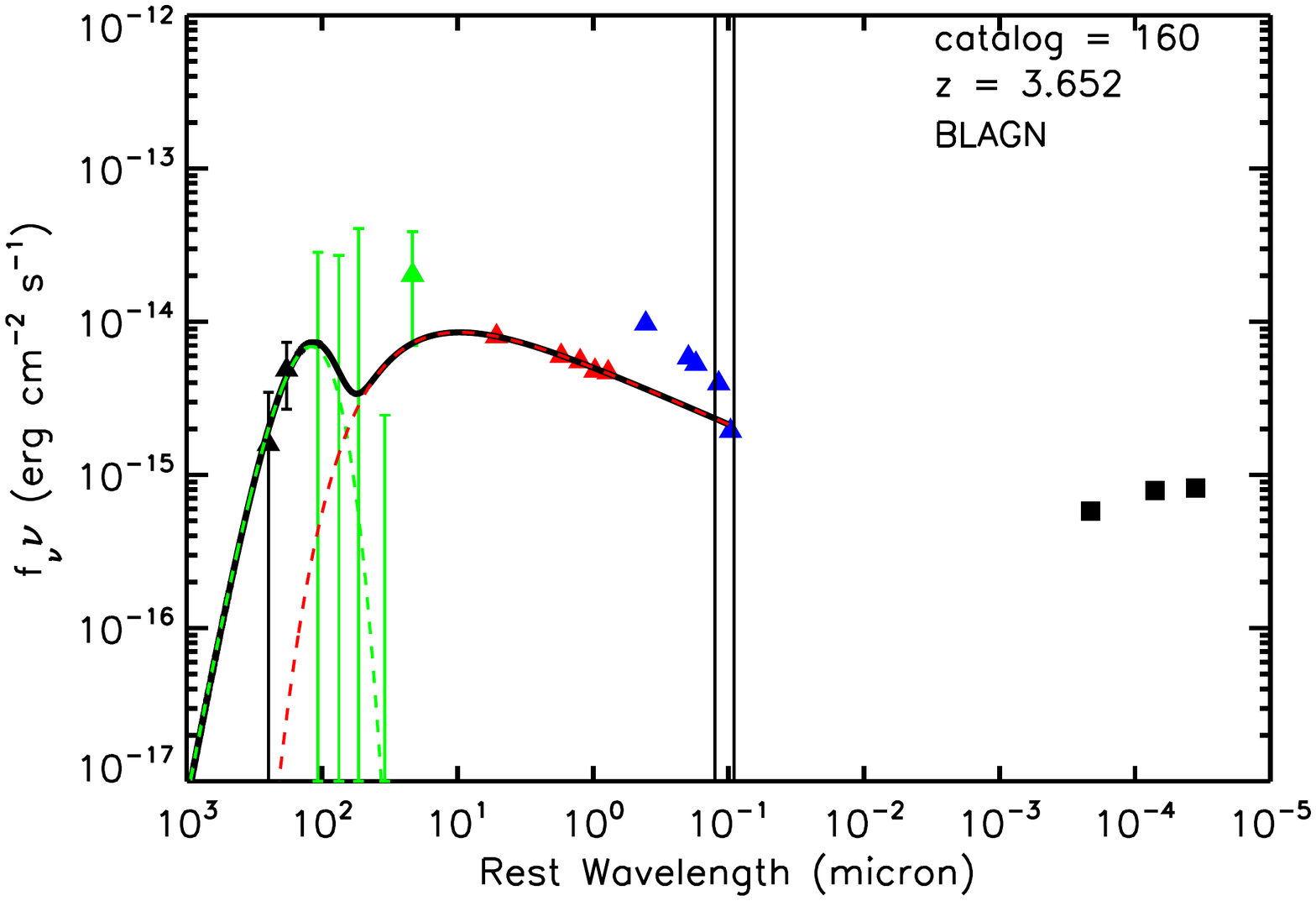}}
\vskip -2.25cm
\centerline{\includegraphics[width=4.25in]{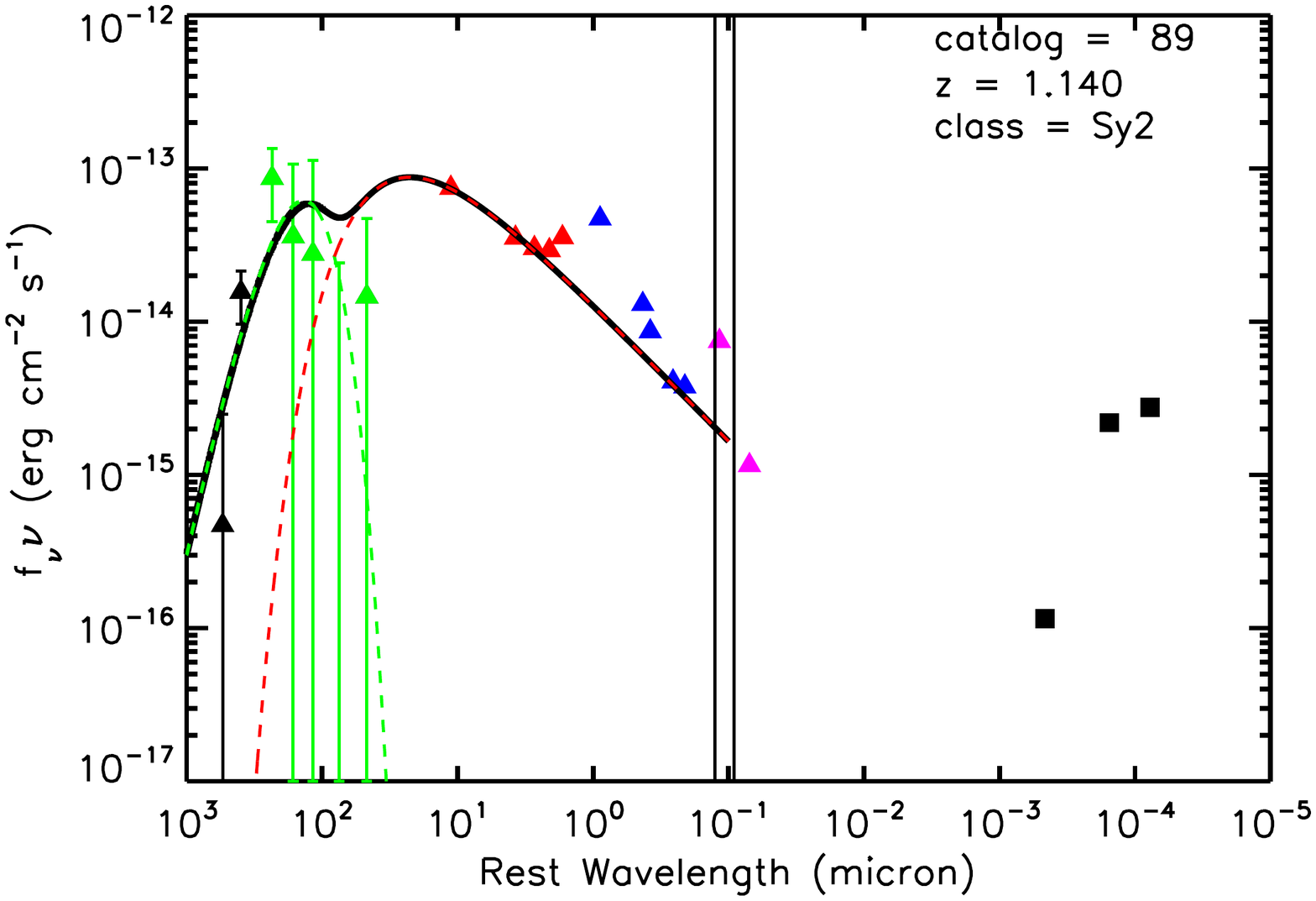}}
\vskip -2.25cm
\centerline{\includegraphics[width=4.25in]{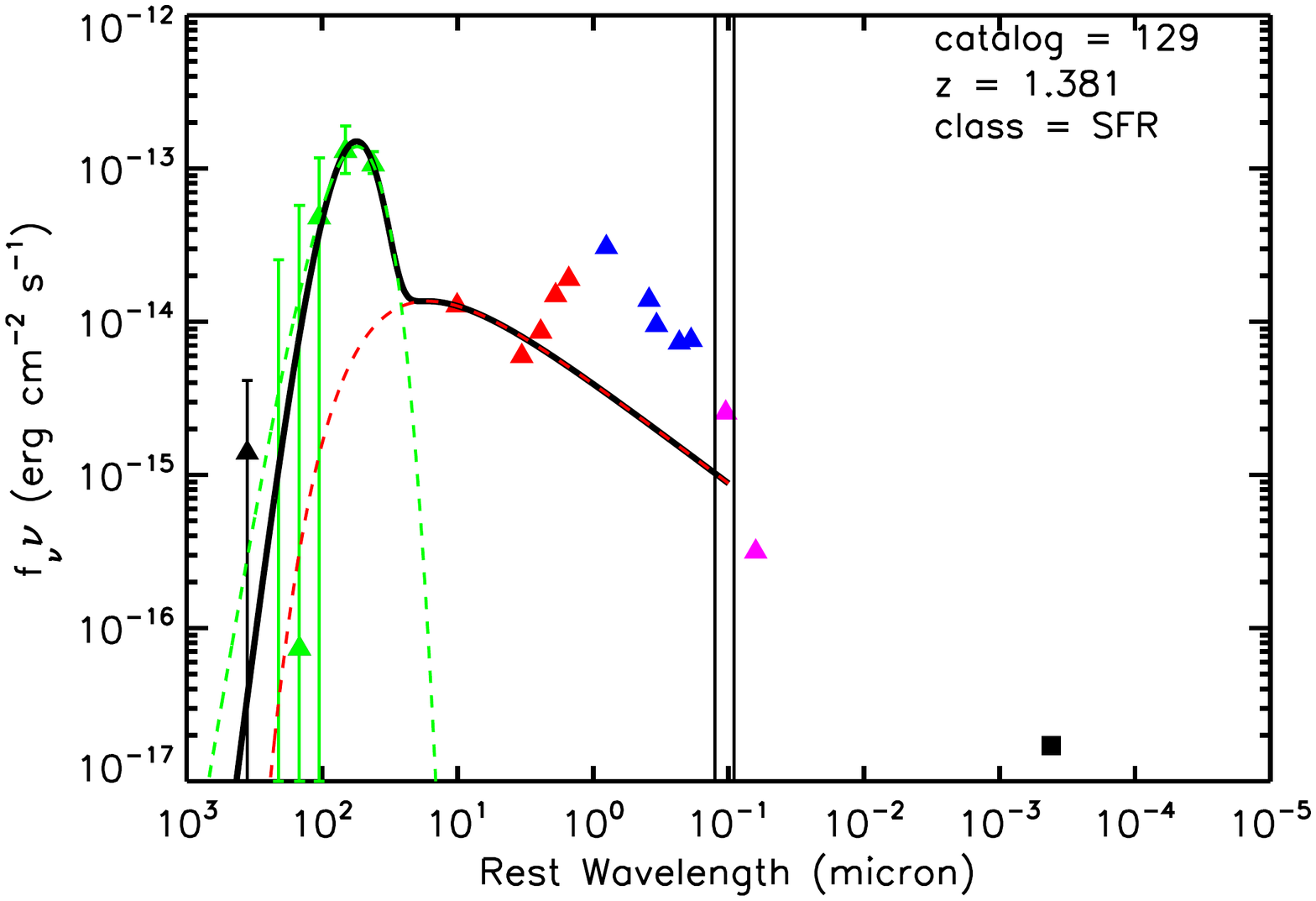}}
\vskip -1.0cm
\caption{
Sample SEDs for (a) a BLAGN, (b) a Sy2, and (c) a star formation dominated galaxy
(black triangles --- ground-based millimeter/submillimeter;
green triangles --- {\em Herschel\/};
red triangles --- {\em Spitzer\/} (IRAC and MIPS);
blue triangles --- some NIR/optical; purple
triangles --- {\em GALEX\/};  black squares --- {\em Chandra\/} 
observed-frame $0.5-2$~keV, $2-4$~keV, and $4-8$~keV).
The long-wavelength data error bars were determined from Monte Carlo measurements 
made at random positions.
The black curve shows the combined gray body and truncated power
law fit, with the individual components shown by the green (gray body)
and the red (truncated power law) dashed curves. 
We do not use any shorter wavelength data than observed-frame 5.6~$\mu$m in the fit.
The black vertical lines show the positions of the Ly$\alpha$ 1216~\AA\ emission line 
and the 912~\AA\ continuum edge.
The catalog numbers given at the top right of each panel are from Alexander et al.\ (2003).
\label{sample_sed}
}
\end{inlinefigure}

We show examples of our SED fits in Figure~\ref{sample_sed} for (a) a BLAGN, (b) a Sy2, and (c) a 
star formation 
dominated galaxy. In each case, we show the individual gray body (green dashed curve), the 
truncated power law (red dashed curve), and the combined fit (black curve). 
Hereafter, we will refer to the luminosities corresponding to the gray body fits as the gray body 
luminosities and those corresponding to the truncated MIR power law fits
integrated above a rest wavelength of $4~\mu$m
as the MIR luminosities.  Note that these quoted luminosities are the total of each component 
(i.e., integrated over all wavelengths).

We can see that the fits in the top two panels of Figure~\ref{sample_sed} are primarily 
constrained by the SCUBA-2 850~$\mu$m data (black triangle).
Indeed, the large error bars on the {\em Herschel\/} data (green triangles)
make it difficult to do SED fits on many of the higher-redshift sources without the addition of the 
submillimeter data.
(Note that the green $70~\mu$m point in Figure~\ref{sample_sed}(a) corresponds to $15~\mu$m at the
galaxy redshift of $z=3.652$ and hence is not expected to be fit by the gray body.)

To avoid cluttering the figure too much, and because the only quantities we are getting from the fits 
are the gray body and MIR luminosities, we do not show the uncertainties on the fits in Figure~\ref{sample_sed}.
However, the uncertainties on the luminosities are determined by the fits, and we show the gray
body luminosity uncertainties on all subsequent figures.

We show full SEDs in Figure~\ref{sample_sed} rather than just the fitted regions to illustrate a few points.
First, as noted in Section~\ref{xraysamp}, the shape of the X-ray data agrees well with the optical spectral 
class of the AGN (i.e., we see a flat X-ray spectral shape for the BLAGN, a drop-off at soft X-rays due to 
obscuration for the Sy2, and a low X-ray flux for the star formation dominated galaxy). 
Second, there is an excess in the NIR/optical over the fit due to the underlying
stellar contributions from the galaxy. This is least pronounced in the BLAGN SED. Finally, at wavelengths shorter
than the Ly$\alpha$ 1216~\AA\ emission line and 912~\AA\ continuum edge (black vertical lines), 
we no longer expect to detect the high-redshift sources.

\section{X-ray Luminosity Dependence in the FIR-Radio Correlation}
\label{xrayFIR-radio}

In the CDF-N, where we have ultradeep radio data, we can construct the FIR-radio correlation,
which has been shown to hold for both star-forming galaxies and
radio-quiet AGNs (e.g., Condon 1992; Mori{\'c} et al.\ 2010).
Indeed, the existence of this correlation has been used as an argument for the FIR luminosity
being primarily produced by star formation, even when the galaxy hosts an AGN
(e.g., Netzer et al.\ 2007).

The radio power is often used as a SFR measure to compare
with other diagnostics (e.g., Cram et al.\ 1998; Hopkins et al.\ 2003; Mushotzky et al.\ 2014). However,
many of the more powerful radio sources are AGN dominated
rather than star formation dominated (e.g., Condon et al.\ 1998; Cowie et al.\ 2004a;
Best et al.\ 2005; Mauch \& Sadler 2007), making this a difficult measure to use unless you
already know which sources are which.

The tight correlation between radio power and FIR
luminosity in low-redshift galaxies is usually parameterized by the
quantity $q$ (e.g., Helou et al.\ 1985; Condon et al.\ 1991),
which is defined as
\begin{equation}
q = \log \left(\frac{L_{\rm FIR}}{3.75\times 10^{12}~{\rm erg~s^{-1}}} \right) - \log \left(\frac{P_{\rm 1.4~GHz}}{\rm erg~s^{-1}~Hz^{-1}} \right) \,,
\end{equation}
where $L_{\rm FIR}$ is the FIR luminosity and
$P_{\rm 1.4~GHz}$ is the rest-frame 1.4~GHz power,
\begin{equation}
P_{1.4~{\rm GHz}}=4\pi {d_L}^2 S_{1.4~{\rm GHz}} 10^{-29}
(1+z)^{\alpha - 1}~{\rm erg~s^{-1}~Hz^{-1}} \,.
\label{eqradio}
\end{equation}
Here $d_L$ is the luminosity distance (cm) and $S_{\rm 1.4~GHz}$
is the 1.4~GHz flux density ($\mu$Jy).
We compute the rest-frame radio power assuming $S_\nu\propto \nu^\alpha$
and a radio spectral index of $\alpha=-0.8$
(Condon 1992; Ibar et al.\ 2010).
The choice of $\alpha$ may not be as appropriate for AGNs as it is for
star formation dominated galaxies, but we adopt a single value for consistency.

Although it is common in the literature to quote FIR luminosities measured over
the broad rest-frame wavelength range $8-1000$~{\micron}
(e.g., Kennicutt et al.\ 1998; Bell 2003; Ivison et al.\ 2010; Mullaney et al.\ 2011), we
avoid this in our analysis, because the $8-1000$~{\micron} definition
covers portions of the spectrum that are likely to be dominated by emission from
the AGN and torus rather than by emission from star formation
(e.g., Horst et al.\ 2008; Gandhi et al.\ 2009; Ichikawa et al.\ 2012).
We instead adopt our gray body luminosities (Section~\ref{indivfits})
for the FIR luminosities of the sources.

\vskip 1.0cm
\begin{inlinefigure}
\centerline{\includegraphics[width=3.25in]{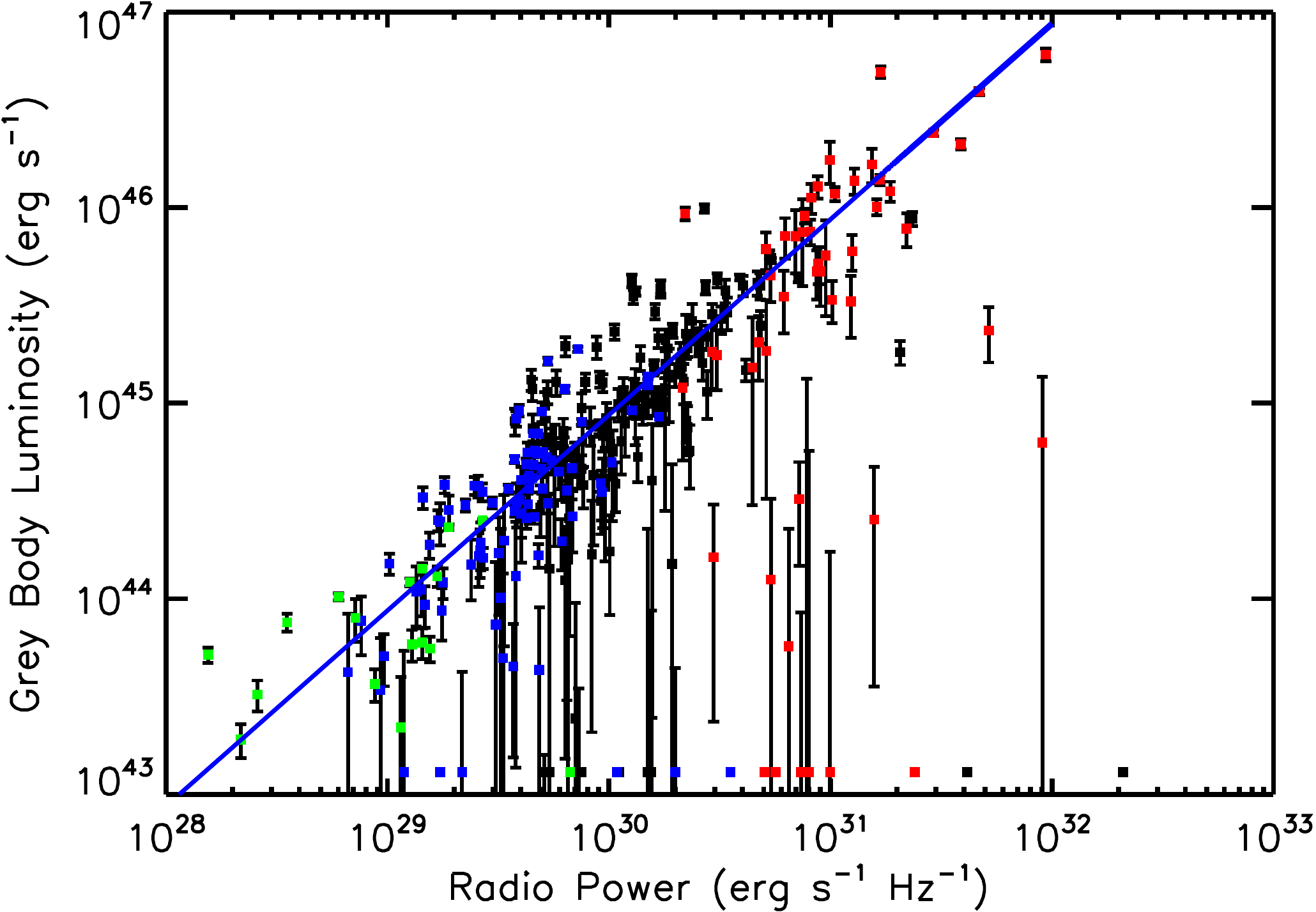}}
\caption{
Gray body luminosity vs. radio power for the sources in our CDF-N radio sample with spectroscopic
or photometric redshifts, excluding any sources with observed-frame $4-8$~keV counterparts
(red --- $z=1.6-4$; black --- $z=0.8-1.6$; blue --- $z=0.4-0.8$; green --- $z=0.2-0.4$).
The radio sources with gray body luminosities below the axis are plotted without
error bars at a nominal $y$ value along the bottom of the plot.
The blue line shows the linear relation for $q=2.36$, which holds over the
full redshift and luminosity range.
\label{radio_radpower_lfir}
}
\end{inlinefigure}

In Figure~\ref{radio_radpower_lfir},
we plot gray body luminosity versus radio power for the sources in our radio
sample with spectroscopic or photometric redshifts
that are not also in our X-ray sample. 
Recent work (Barger et al.\ 2012, 2014; Thomson et al.\ 2014) 
has shown that the FIR-radio correlation
holds to very high redshifts ($z\sim5$), at least for ULIRGs,
and the present work confirms this. 
Approximately 90$\%$ of the radio sources follow a tight FIR-radio correlation over
the wide redshift and gray body luminosity ranges of the sample. 
Including all of the sources in the plot, 
we find $q=2.36\pm0.01$, while if we restrict to sources 
at $z=1.6-4$, then we find $q=2.23\pm0.05$. 
The blue line shows the linear relation for $q=2.36$.

We can now test how well the gray body luminosities for our $z=0.8-4$ X-ray
sample (including sources with only upper limits on their radio power) follow the 
FIR-radio correlation. In Figure~\ref{radpower_lfir}, we plot gray body luminosity 
versus radio power (symbols are color-coded 
by spectral type), separated according to the definition of an X-ray quasar:
(a) $L_X>5 \times 10^{43}$~erg~s$^{-1}$ and 
(b) $L_X\le 5 \times 10^{43}$~erg~s$^{-1}$.
While the X-ray quasars cover a relatively wide range 
of gray body luminosities, there are many low values.
In contrast, the X-ray less luminous AGNs mostly obey the correlation. 
(As an aside, we note that the observed 
dependence on X-ray luminosity does not appear to depend on the spectral type.)

\begin{inlinefigure}
\includegraphics[width=3.75in]{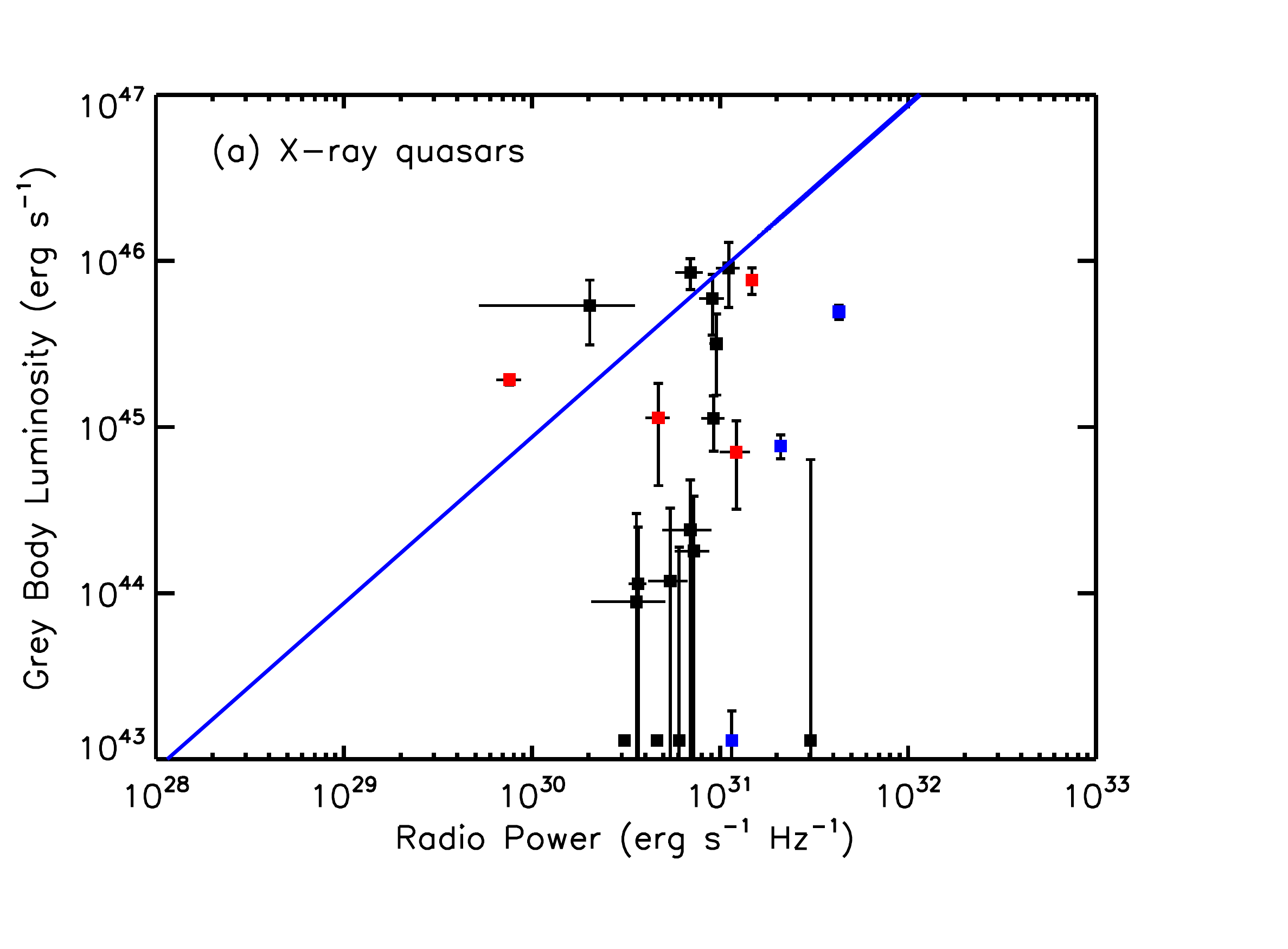}
\vskip -1.0cm
\includegraphics[width=3.75in]{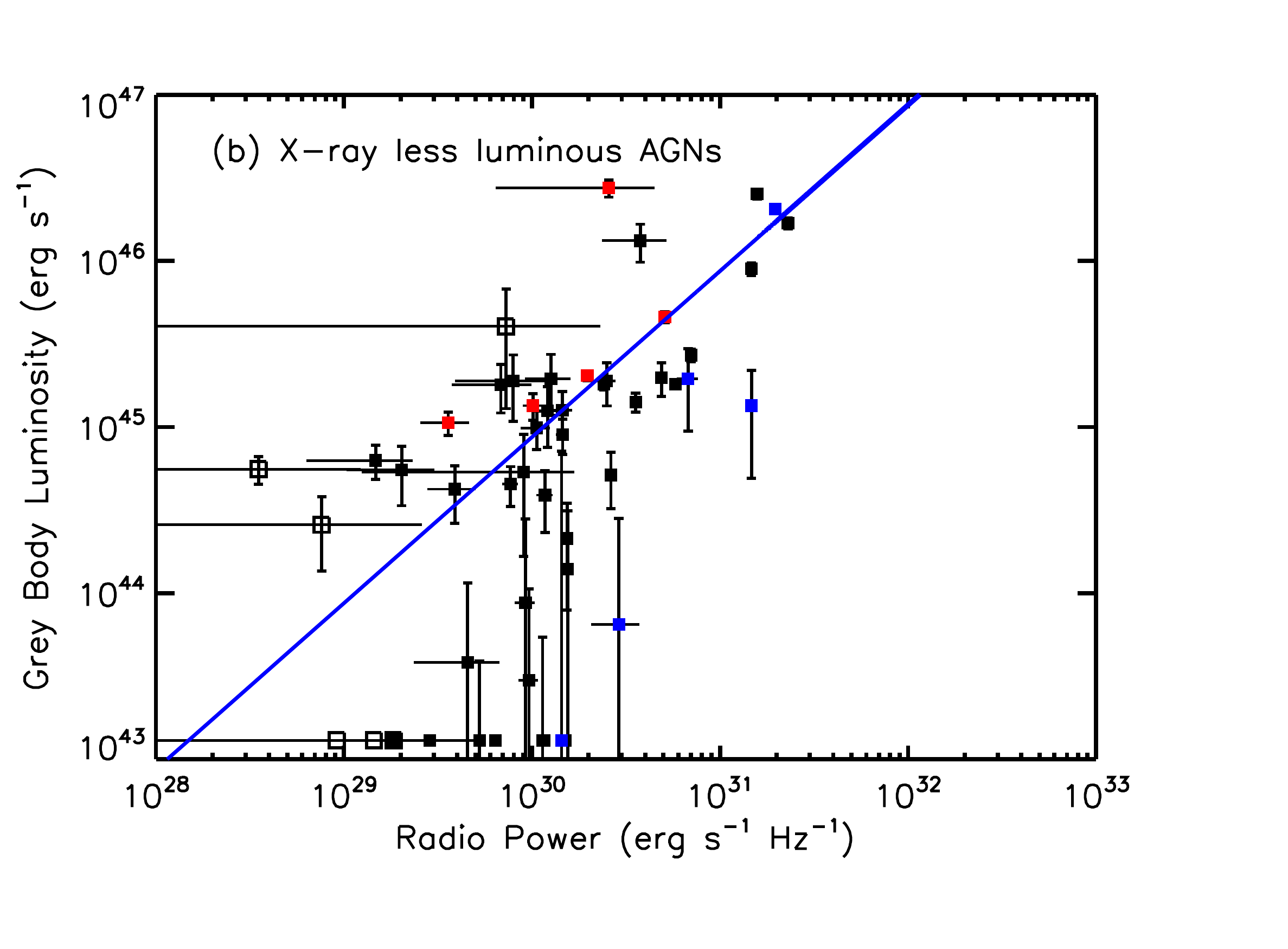}
\vskip -0.5cm
\caption{
Gray body luminosity vs. radio power for our CDF-N X-ray sample at $z=0.8-4$ 
(red squares --- BLAGNs; blue squares --- Sy2s; black squares --- other; 
open squares --- sources with only 
upper limits on their radio power), separated by X-ray luminosity:
(a) $L_X>5 \times 10^{43}$~erg~s$^{-1}$ and
(b) $L_X\le5 \times 10^{43}$~erg~s$^{-1}$. 
The X-ray sources with gray body luminosities below the axis are plotted without vertical
error bars at a nominal $y$ value along the bottom of each panel.
In each panel, the blue diagonal line shows the 
FIR-radio correlation from Figure~\ref{radio_radpower_lfir}.
\label{radpower_lfir}
}
\end{inlinefigure}

Figure~\ref{radpower_lfir} therefore suggests that for X-ray quasars, the radio power is 
not related to the star formation in the host galaxies and must instead be dominated by the AGN. 
However, for the X-ray less luminous AGNs, the radio power is consistent with the 
FIR-radio correlation and therefore is probably dominated by the star formation in the
host galaxies.

Since luminosity-luminosity plots constructed from flux-limited samples can produce 
apparent correlations that are not real, it is also important to examine ratios.
In Figure~\ref{radpower-rat}, we plot the ratio of gray body luminosity to radio power 
versus X-ray luminosity for the X-ray sources with
$P_{\rm 1.4~GHz}>10^{30}$~erg~s$^{-1}$~Hz$^{-1}$.
We also plot the distribution of the radio sample (using the
same radio power threshold) that are not also in the X-ray sample 
in histogram form (blue) and their mean value
(blue horizontal line). Consistent with Figure~\ref{radpower_lfir}, 
the ratio drops at high X-ray luminosities. Moreover, the transition is occurring slightly 
above the X-ray quasar luminosity definition (red vertical line).
Thus, X-ray luminosity can be used as one diagnostic for determining when SFRs
may be estimated from radio power.

\vskip 0.25cm
\begin{inlinefigure}
\includegraphics[width=3.8in]{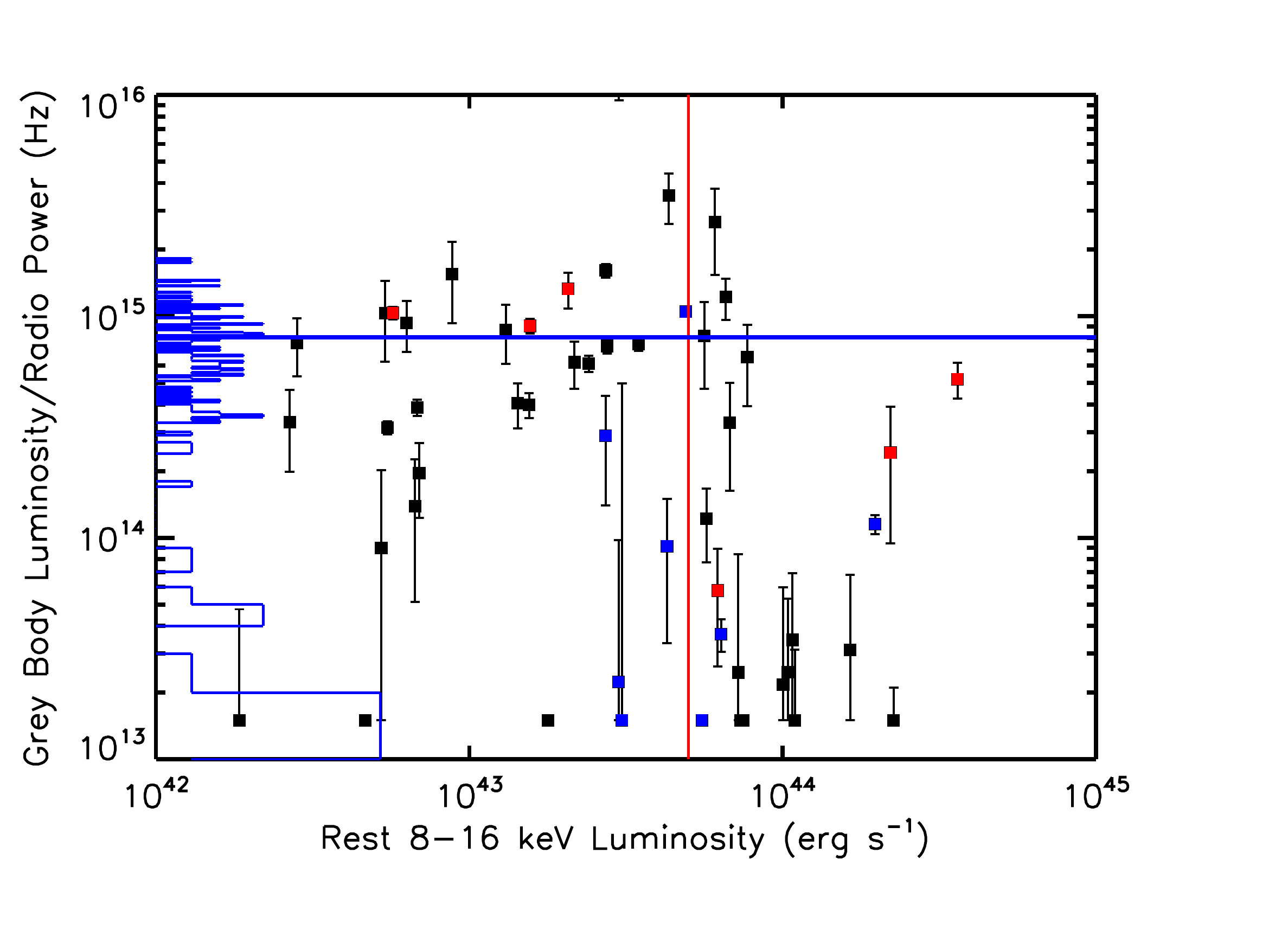}
\vskip -0.5cm
\caption{
Ratio of gray body luminosity to radio power vs. $L_X$ 
for our CDF-N X-ray sample
at $z=0.8-4$ (red squares --- BLAGNs; blue squares --- Sy2s; 
black squares --- other) having $P_{\rm 1.4~GHz}>10^{30}$~erg~s$^{-1}$~Hz$^{-1}$.
The blue histogram shows the distribution of our CDF-N radio sample at $z=0.8-4$ (using the same radio
power threshold as above), excluding any sources with observed-frame $4-8$~keV counterparts,
and the blue horizontal line shows their mean value.
The X-ray and radio sources with ratios below the axis are plotted at a nominal $y$ value along 
the bottom of the plot. The red vertical line shows the X-ray quasar definition.
\label{radpower-rat}
}
\end{inlinefigure}

\section{X-ray Luminosity Dependence in the FIR-MIR Correlation}
\label{xrayFIR-MIR}

We now use the larger combined CDF-N+CDF-S X-ray sample
to investigate the relative strengths of the FIR and MIR components of the X-ray sources on 
an individual basis. 
In Figure~\ref{radpowercdfs_lmir}, we plot gray body luminosity versus MIR luminosity 
for our $z=0.8-4$ radio sample that are not also in the X-ray sample (blue dots).
We also show the linear relation obtained
from these points (blue line). We can see that the radio sources follow the relation
tightly, confirming the existence of a MIR-radio correlation in star formation dominated galaxies.

We also show on the figure the combined CDF-N and CDF-S X-ray sample at $z=0.8-4$ that have
(a) $L_X>5 \times 10^{43}$~erg~s$^{-1}$ and 
(b) $L_X\le 5 \times 10^{43}$~erg~s$^{-1}$ (colored squares).
While most of the X-ray less luminous sources follow the correlation, suggesting that 
these sources have significant FIR emission due to star formation in the host galaxies,
most of the X-ray quasars have gray body luminosities that are low compared to the 
correlation, as would be expected if there was little star formation.
(Note that the MIR luminosities stay high in the X-ray quasars due to the AGN contribution.)

We again wish to check that such a luminosity-luminosity plot constructed from flux-limited samples 
is not producing apparent correlations that are not real. Thus, 
in Figure~\ref{mirfir-rat}, we show the ratio of gray body luminosity to MIR luminosity 
versus X-ray luminosity for the X-ray sources with
$P_{\rm 1.4~GHz}>10^{30}$~erg~s$^{-1}$~Hz$^{-1}$. 
We also plot in histogram form (blue) the distribution of the radio sample (using the
same radio power threshold) that are not also in the X-ray sample and their mean value
(blue horizontal line).

\vskip 0.5cm
\begin{inlinefigure}
\centerline{\includegraphics[width=3.25in]{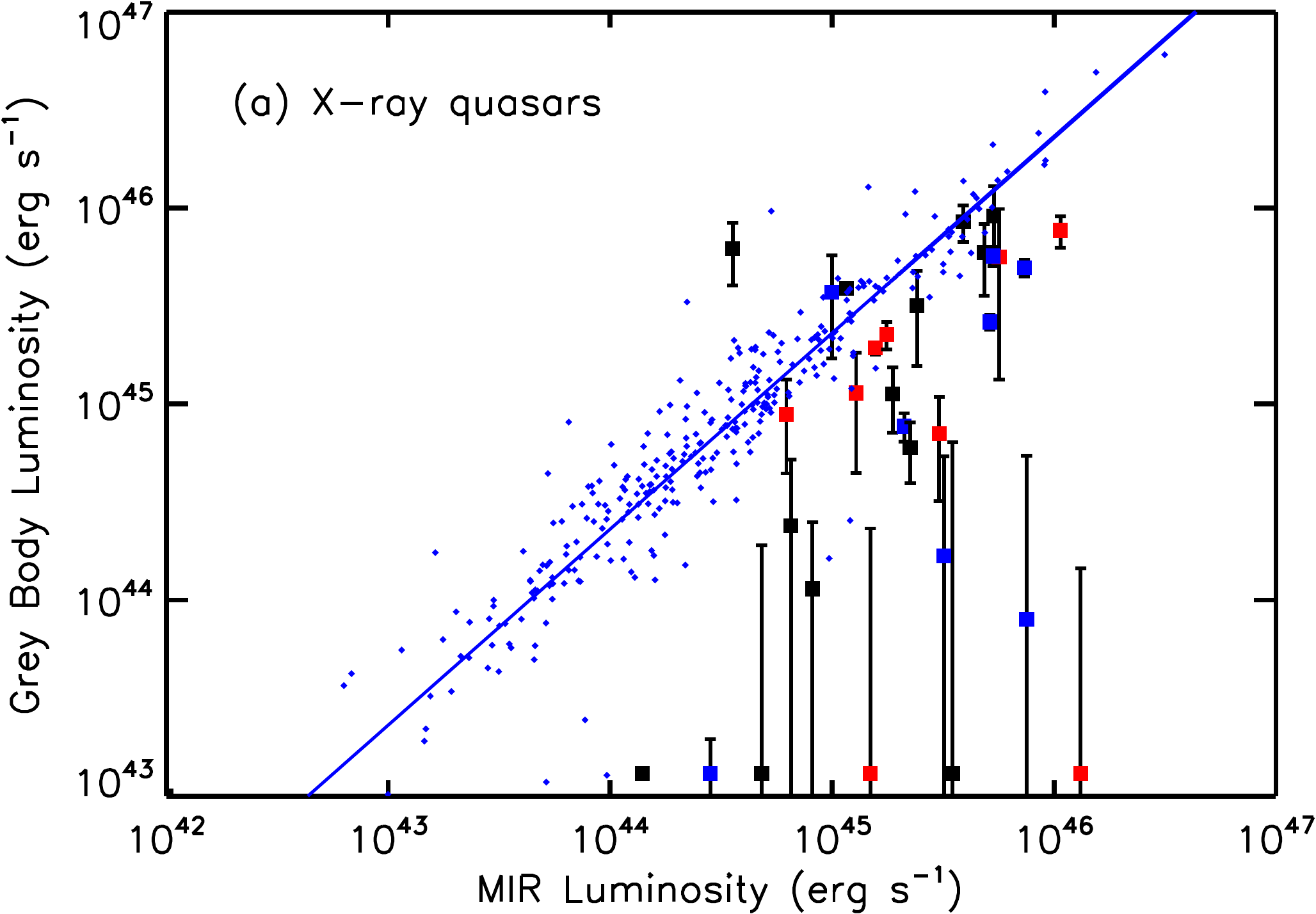}}
\vskip 0.5cm
\centerline{\includegraphics[width=3.25in]{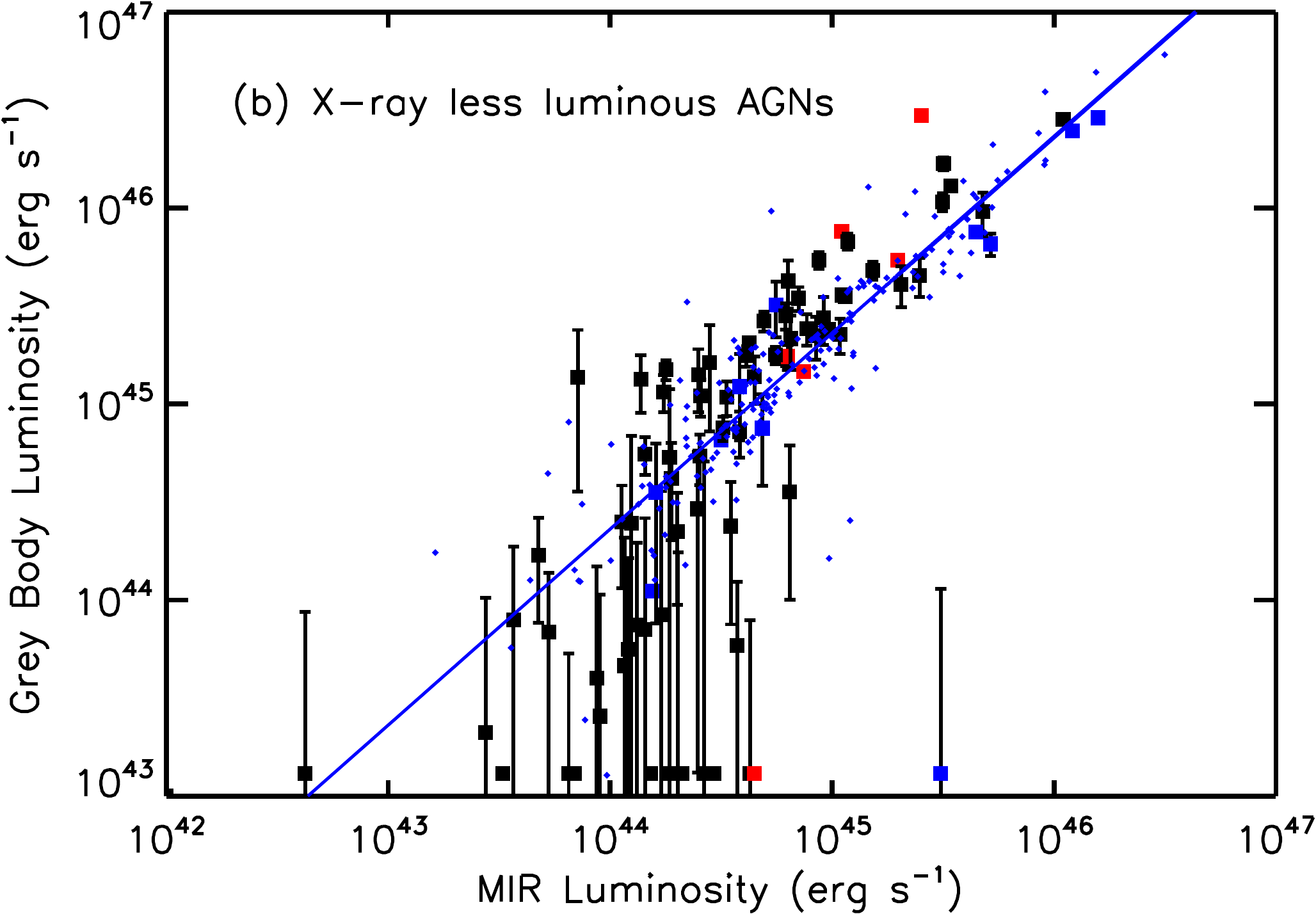}}
\caption{
Gray body luminosity vs. MIR luminosity for our combined CDF-N and CDF-S
X-ray sample at $z=0.8-4$ (red squares --- BLAGNs;
blue squares --- Sy2s; black squares --- other),
separated by X-ray luminosity:
(a) $L_X>5 \times 10^{43}$~erg~s$^{-1}$ and
(b) $L_X\le5 \times 10^{43}$~erg~s$^{-1}$. 
The blue dots show the sources in our CDF-N radio sample at $z=0.8-4$, 
excluding any sources with observed-frame $4-8$~keV counterparts.
The X-ray and radio sources with gray body luminosities below the axis are plotted 
without vertical error bars at a nominal $y$ value along the bottom of each panel.
The blue line shows the linear relation determined from the radio sources.
\label{radpowercdfs_lmir}
}
\end{inlinefigure}

We see an abrupt drop in the gray body to MIR luminosity ratios for the most X-ray luminous 
sources, with the transition X-ray luminosity lying slightly above the X-ray quasar definition 
(red vertical line). In contrast, the ratios for the X-ray less luminous sources remain relatively 
constant (also apparent in the blue histogram showing the radio sample). 
We interpret this as evidence that there is little star formation taking place in the hosts of X-ray 
luminous AGNs, confirming what we saw schematically in Figure~\ref{stack_sed_lx}.

The above results alternatively could be ascribed to the
superpositions of star-forming galaxy SEDs and varying AGN SEDs.
However, we suspect that this might lead to a less abrupt evolution 
of the gray body to MIR luminosity ratios to high X-ray luminosities.

\vskip -0.8cm
\begin{inlinefigure}
\hskip -0.5cm
\centerline{\includegraphics[width=4.5in]{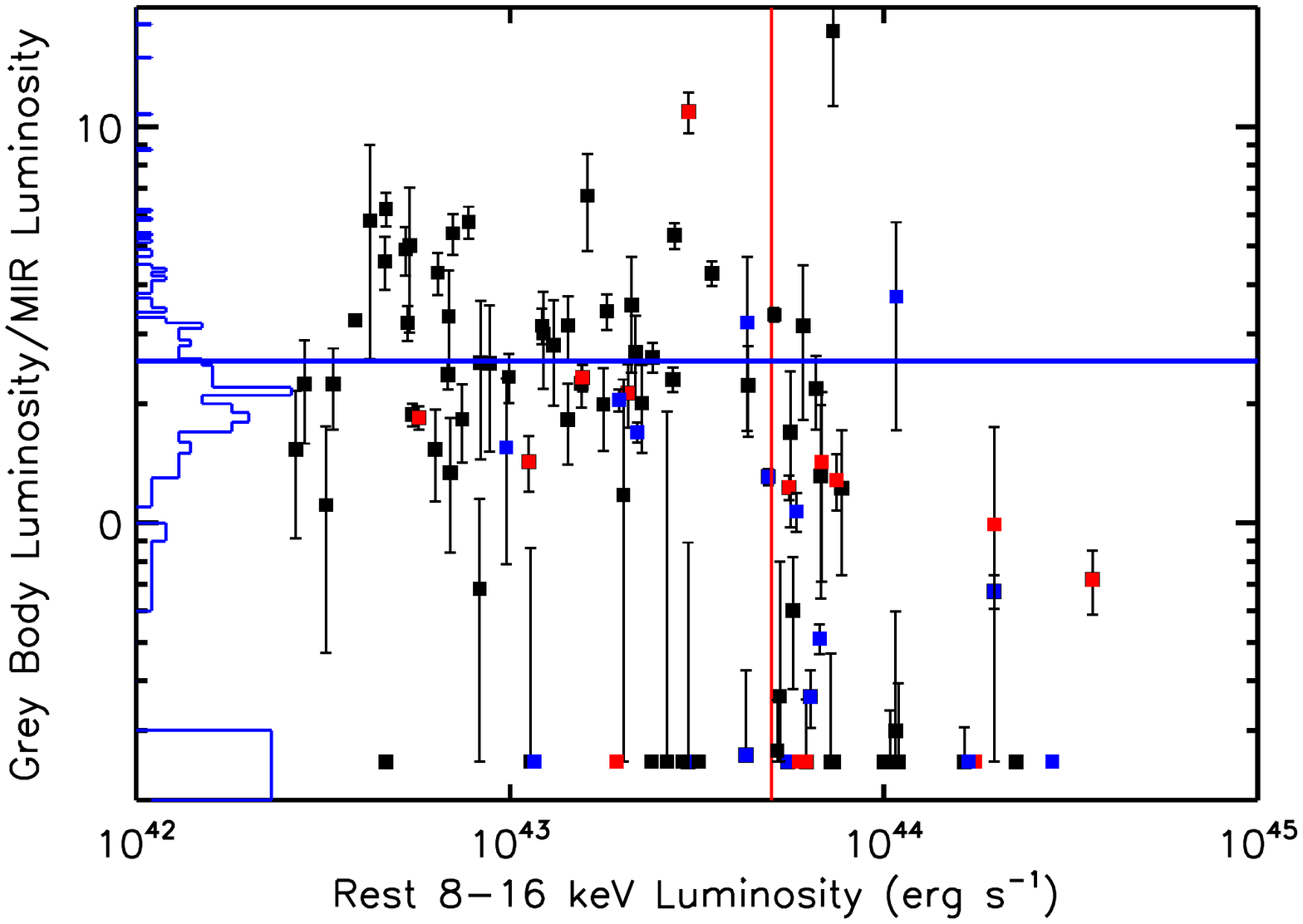}}
\vskip -1.25cm
\caption{
Ratio of gray body luminosity to MIR luminosity vs. $L_X$ for our combined CDF-N and
CDF-S X-ray sample at $z=0.8-4$ (red squares --- BLAGNs; blue squares --- Sy2s; 
black squares --- other) having $P_{\rm 1.4~GHz}>10^{30}$~erg~s$^{-1}$~Hz$^{-1}$.
The blue histogram shows the distribution of our CDF-N radio sample at $z=0.8-4$ 
(using the same radio power threshold as above), excluding any sources with observed-frame 
$4-8$~keV counterparts, and the blue horizontal line shows their mean value.
The X-ray and radio sources with ratios below the axis are plotted at a nominal $y$ value along 
the bottom of the plot. The red vertical line shows the X-ray quasar definition.
\label{mirfir-rat}
}
\end{inlinefigure}

\section{Luminosity Distributions}
\label{lumdist}

We finally turn to the distribution of gray body luminosities, $L_{Gray}$, as a function of $L_X$.
In Figure~\ref{histdist}(a), we show this distribution
for the $z=1.5-4.5$ sources in the CDF-N and CDF-S fields, divided into
two intervals of $\log L_X$:  $42-44$ and $>44$~erg~s$^{-1}$. 

In order to include the COSMOS field, we also computed a simpler set of
luminosities, $L_{\rm Arp}$, which are based on the Arp~220 SED and only the measured
submillimeter flux and redshift instead of on the combined 
FIR gray body and MIR power law fits.
In Figure~\ref{histdist}(b),
we show this luminosity distribution for the same redshift and $\log L_X$ intervals 
as in Figure~\ref{histdist}(a). 
The larger uncertainties in these luminosities cause a larger spread in the
distributions, but we now have high enough numbers to run a two tailed Mann-Whitney test.
We find only a 0.026 probability that the two distributions are consistent with one another, 
which is similar to our earlier results.
We also show on the figures the expected distributions based on the mean errors if there were 
no signal in the sources (black curves). While the $\log L_X>44$~erg~s$^{-1}$ source distribution is
consistent with no signal, the $\log L_X=42-44$~erg~s$^{-1}$ source distribution is offset
due to signal.

In Figure~\ref{histdist}(a), the mean values of the histograms (green dashed vertical lines) are 
$L_{Gray}=3.4\times10^{45}$~erg~s$^{-1}$ for the low $L_X$ interval and 
$L_{Gray}=1.6\times10^{45}$~erg~s$^{-1}$ for the high $L_X$ interval,
which would place the hosts in the near-ULIRG category 
(a ULIRG corresponds to $L_{FIR}>3.8\times 10^{45}$~erg~s$^{-1}$). 

However, as can be seen from all the components of Figure~\ref{histdist}, the distributions 
are highly skewed, resulting in the mean values being dominated by a small number 
of high-luminosity galaxies. In the blue upper histogram of Figure~\ref{histdist}(a), just 
6 out of 67 galaxies contain half the light and dominate the determination of the mean. 
Thus, simple stacking or averaging analyses cannot adequately describe
the behavior of the  bulk of the galaxies.

\vskip 0.25cm
\begin{inlinefigure}
\includegraphics[width=3.4in,angle=0]{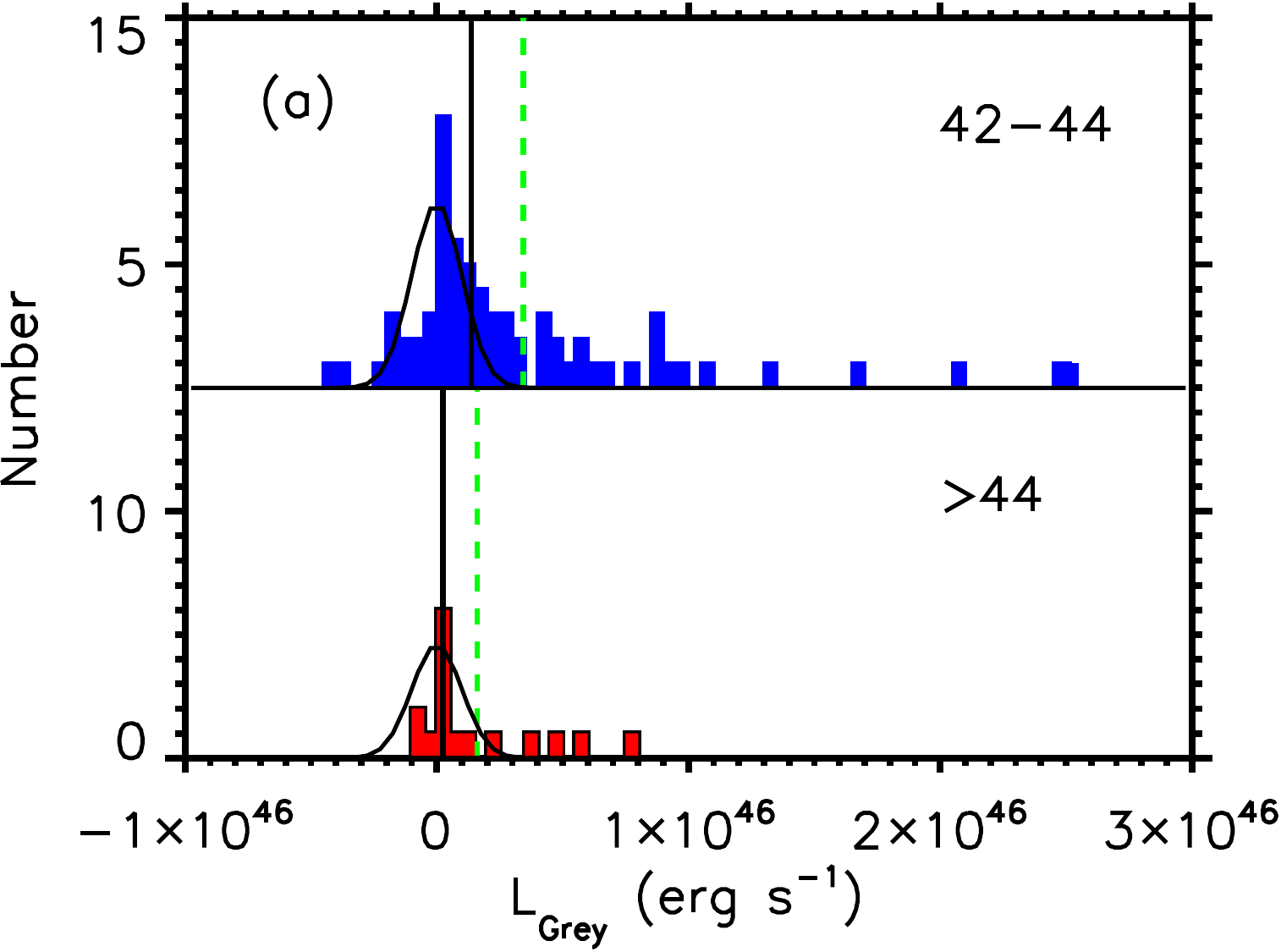}
\vskip 0.5cm
\includegraphics[width=3.4in,angle=0]{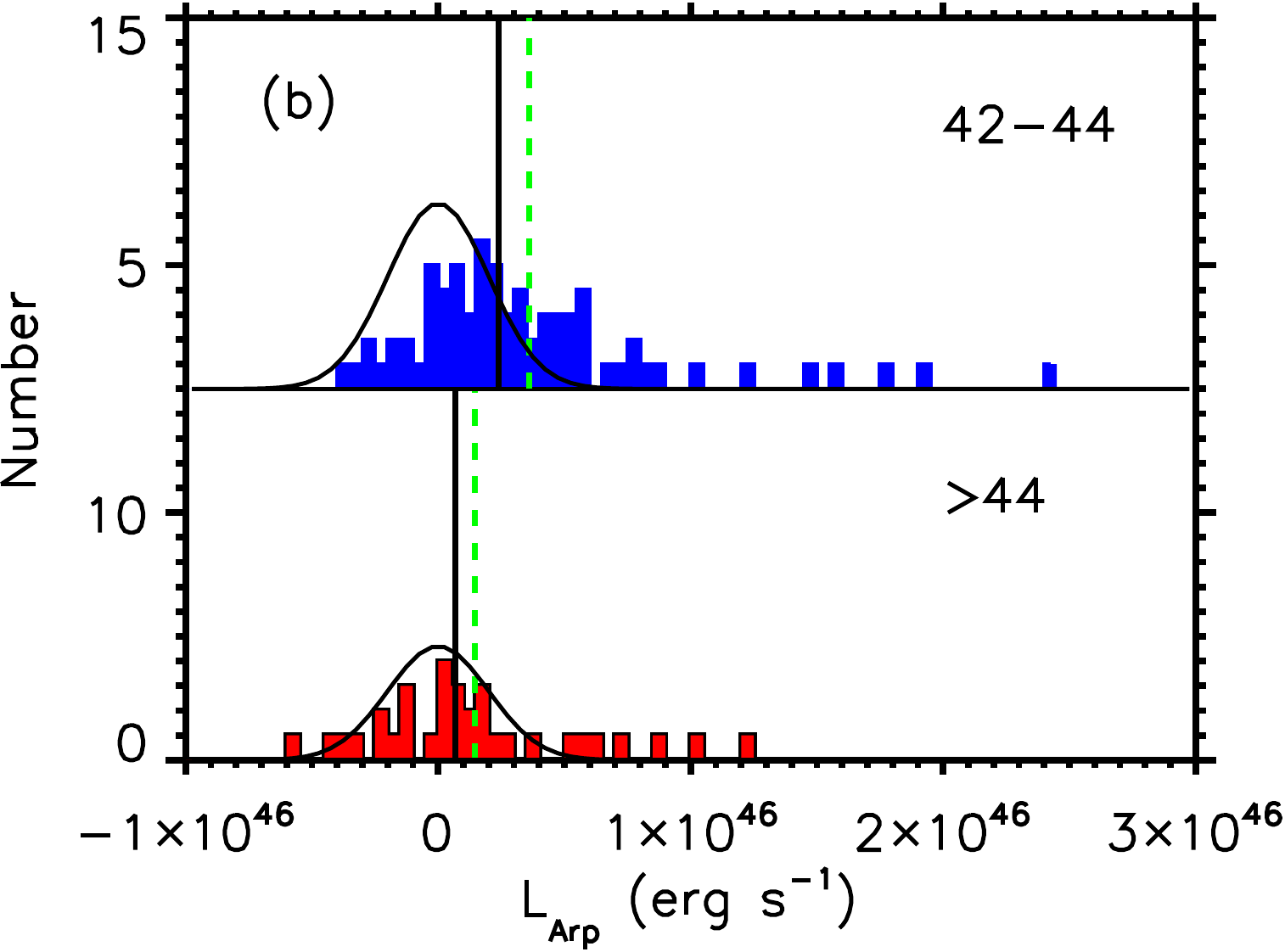}
\caption{
Distribution of FIR luminosities for the X-ray sources with redshifts $z=1.5-4.5$, divided into 
two intervals of $\log L_X$: $42-44$~erg~s$^{-1}$ (blue upper histogram)
and $>44$~erg~s$^{-1}$ (red lower histogram). In (a), we show the gray body luminosities
for the CDF-S and CDF-N fields. In (b), we show the luminosities computed instead
using the Arp~220 SED, which makes it possible to include the COSMOS sources.
The green dashed vertical lines in each panel show the mean values of each histogram, 
and the black vertical lines show the median values. The black curves in each panel 
show the expected distributions based on the mean errors if there were no signal.
\label{histdist}
}
\end{inlinefigure}

The skewness of the distributions could suggest that the duty cycle for the optically
luminous phase of the host galaxies is only about 10\% of the duty cycle of the AGN
themselves, with galaxies mostly being quiescent when the AGN is luminous with
occasional strong starburst episodes.

We may use the median instead to represent more properly the behavior of the typical 
galaxy. In Figure~\ref{histdist}(a), we find median luminosities (black vertical lines) of 
$L_{Gray}=1.4\pm0.8\times10^{45}$~erg~s$^{-1}$ for the low $L_X$ interval
and $L_{Gray}=0.24\, (0.17,0.37)\times10^{45}$~erg~s$^{-1}$ for the high $L_X$ interval.
Here the errors are the 68\% confidence range. These median values show 
that the typical hosts of the low $L_X$ interval sources lie in the LIRG 
range (a LIRG corresponds to $L_{FIR}>3.8\times 10^{44}$~erg~s$^{-1}$), while the typical 
hosts of the high $L_X$ interval sources lie in the sub-LIRG range. Only a very small number 
of X-ray sources lie in very luminous (ULIRG or greater) hosts.
 
\section{Summary}
\label{summary}

In this paper, we examined the amount of dusty star formation taking place in the host galaxies 
of a $4-8$~keV AGN sample in the CDF-N and CDF-S fields using ultradeep 850~$\mu$m 
SCUBA-2 images of both fields and an extremely deep 1.4~GHz VLA image of the CDF-N.
Supplementing this sample with a brighter X-ray sample in the COSMOS field with SCUBA-2
data, we first measured the submillimeter fluxes of the X-ray sources with spectroscopic 
or photometric redshifts $z>1$. 
We found a dependence of 850~$\mu$m flux on X-ray luminosity, with 
the error-weighted means peaking at $L_X=10^{43}-10^{43.5}$~erg~s$^{-1}$ 
before dropping by $4\sigma$ to $L_X=10^{43.5}-10^{44}$~erg~s$^{-1}$, and by an even
larger amount to $L_X=10^{44}-10^{44.5}$~erg~s$^{-1}$.
Monte Carlo simulations give only a 0.013 probability that the 850~$\mu$m fluxes drawn from
the $L_X>10^{44}$~erg~s$^{-1}$ population are as large as those drawn from the 
$L_X=10^{43}-10^{44}$~erg~s$^{-1}$ population.
Assuming the FIR light is produced mostly by star formation, we interpreted this result as 
an initial rise in the host galaxy SFRs with increasing X-ray luminosity
followed by a drop in the SFRs to the highest X-ray luminosities.

Substantially more information is contained in the full SEDs of the X-ray sources. 
Given the extensive multiwavelength data available, including from {\em Herschel},
we constructed average SEDs for our X-ray sample to show schematically the observed
dependence on X-ray luminosity.  We chose three 
redshift ranges ($z=2.0-4.5$; $0.8-2.0$; and $0.4-0.8$) and four X-ray luminosity intervals 
($L_{X}>10^{44}$~erg~s$^{-1}$; $10^{43.5}-10^{44}$;
$10^{43}-10^{43.5}$; $10^{42.0}-10^{43}$).
We saw that the FIR luminosities of the host galaxies rose with increasing redshift. However,
within each redshift range, the average SED of the highest X-ray luminosity 
interval was cut off at long wavelengths. This reinforces the idea that the SFRs 
in the host galaxies of the most X-ray luminous sources at any redshift are too low 
to produce a substantial FIR luminosity; however, to analyze this effect statistically, 
we need to fit the individual source SEDs. 

We performed individual SED fits on both the X-ray and radio sources
using a 5 parameter combined FIR gray body plus truncated MIR power law fit. 
We used the resulting gray body and MIR luminosities in the CDF-N to confirm 
the FIR-radio and MIR-radio correlations for the $z=0.2-4$ radio sources that are not also in the X-ray sample.

We then looked to see whether the $z=0.8-4$ X-ray sample in the CDF-N obeyed the FIR-radio correlation.
We found that most of the X-ray quasars ($L_X>5\times 10^{43}$~ergs~s$^{-1}$) did not,
while most of the X-ray less luminous AGNs did. 
Thus, for the X-ray quasars, the radio power does not
appear to be related to the star formation in the host galaxies, while for the X-ray less luminous AGNs,
the radio power appears to be dominated by the star formation. 
This suggests that X-ray luminosity is useful as a 
diagnostic for determining when radio power may be used to estimate SFRs. 

We next investigated the relative strengths of the FIR and MIR components of the X-ray sources. 
For the combined CDF-N and CDF-S 
$z=0.8-4$ X-ray sample, we found that the X-ray less luminous AGNs generally followed the
FIR-MIR correlation, while most of the X-ray quasars lay below the correlation. 
We interpreted the FIR luminosities as being low in the host galaxies of the X-ray quasars
due to the lack of star 
formation, while the MIR luminosities stayed high due to the AGN contribution.

Finally, we analyzed the distribution of FIR luminosities as a function of X-ray luminosity and
found that the median represents the behavior of the typical galaxy better than the mean, which
is skewed by a small number of sources ($\sim10$\% of the sample).
Thus, stacking or averaging analyses overestimate the level of star formation taking place
in the bulk of the X-ray sample, and analyses of individual sources, such as those presented 
in this paper, are needed.

\acknowledgements

We thank B.~Lehmer for providing the $4-8$~keV catalog in the CDF-S and G.~Wilson 
for providing the 1.2~mm images in the CDF-N. We would also like to
thank the referee for a very interesting and thought provoking report that helped us to
improve the manuscript and Richard Mushotzky and Joe Silk for thoughtful comments on an 
earlier draft of the paper. We gratefully acknowledge support from 
the David and Lucile Packard Foundation (A.~J.~B.)
and NSF grants AST-1313150 (A.~J.~B.) and AST-1313309 (L.~L.~C.).
C.-C.~C. acknowledges support from the ERC Advanced Investigator programme
DUSTYGAL 321334.
We acknowledge the cultural significance that the summit of 
Mauna Kea has to the indigenous Hawaiian community.



\begin{references}

\reference{alex03b}
Alexander, D. M., Bauer, F. E., Brandt, W. N., et al.\ 2003, \aj, 126, 539 

\reference{balestra10}
Balestra, I., Mainieri, V., Popesso, P., et al.\ 2010, A\&A, 512, 12

\reference{barger05}
Barger, A. J., Cowie, L. L., Mushotzky, R. F., et al.\ 2005, \aj, 129, 578

\reference{barger07} 
Barger, A. J., Cowie, L. L., \& Wang, W.-H.\ 2007, \apj, 654, 764

\reference{barger08}
Barger, A. J., Cowie, L. L., \& Wang, W.-H.\ 2008, \apj, 689, 687

\reference{barger12} 
Barger, A. J., Wang, W.-H., Cowie, L. L., et al.\ 2012, \apj, 761, 89

\reference{barger14}
Barger, A. J., Cowie, L. L., Chen, C.-C., et al.\ 2014, \apj, 784, 9

\reference{bauer04} 
Bauer, F. E., Alexander, D. M., Brandt, W. N., et al.\ 2004, \aj, 128, 2048

\reference{bell03} 
Bell, E. F.\ 2003, \apj, 586, 794

\reference{best05}
Best, P. N., Kauffmann, G., Heckman, T. M., \& Ivezi{\'c}, \v{Z}.\ 2005,
\mnras, 362, 9

\reference{bower11} 
Bower, R. G., Benson, A. J., Malbon, R., et al.\ 2006, \mnras, 370, 645

\reference{cano12}
Cano-D{\'i}az, M., Maiolino, R., Marconi, A., et al.\ 2012, A\&A, 537, L8

\reference{capak04} 
Capak, P., Cowie, L. L., Hu, E. M., et al.\ 2004, \aj, 127, 180

\reference{casey12} 
Casey, C. M.\ 2012, \mnras, 425, 3094

\reference{caseyetal12}
Casey, C. M., Berta, S., B{\'e}thermin, M., et al.\ 2012, \apj, 761, 140

\reference{casey13}
Casey, C. M., Chen, C.-C., Cowie, L. L., et al.\ 2013, \mnras, 436, 1919

\reference{chapin13}
Chapin, E. L., Berry, D. S., Gibb, A G., et al.\ 2013, \mnras, 430, 2545

\reference{chary01} 
Chary, R., \& Elbaz, D.\ 2001, \apj, 556, 562

\reference{chen13a}
Chen, C.-C., Cowie, L. L., Barger, A. J., et al.\ 2013a, \apj, 762, 81 

\reference{chen13b} 
Chen, C.-C., Cowie, L. L., Barger, A. J., et al.\ 2013b, \apj, 776, 131 

\reference{chenhickox13}
Chen, C.-T. J., Hickox, R. C., Alberts, S., et al.\ 2013, \apj, 773, 3

\reference{civano12}
Civano, F., Elvis, M., Brusa, M., et al.\ 2012, \apjs, 201, 30

\reference{cohen00}
Cohen, J. G., Hogg, D. W., Blandford, R., et al.\ 2000, \apj, 538, 29

\reference{condon92} 
Condon, J. J.\ 1992, ARA\&A, 30, 575

\reference{condon91} 
Condon, J. J., Anderson, M. L., \& Helou, G.\ 1991, \apj, 376, 95

\reference{condon98}
Condon, J. J., Yin, Q. F., Thuan, T. X., \& Boller, T.\ 1998, \aj, 116, 2682

\reference{cooper11}
Cooper, M. C., Aird, J. A., Coil, A. L., et al.\ 2011, \apjs, 193, 14

\reference{cooper12}
Cooper, M. C., Yan, R., Dickinson, M., et al.\ 2012, \mnras, 425, 2116

\reference{cowie15}
Cowie, L. L., et al.\ 2015, \apj, in preparation

\reference{cowie03} 
Cowie, L. L., Barger, A. J., Bautz, M. W., Brandt, W. N., \& Garmire, G. P.\ 2003, \apj, 584, L57

\reference{cowie12}
Cowie, L. L., Barger, A. J., \& Hasinger, G.\ 2012, \apj, 748, 50

\reference{cowie04a}
Cowie, L. L., Barger, A. J., Fomalont, E. B., \& Capak, P.\ 2004a, \apj, 603, L69

\reference{cowie04b}
Cowie, L. L., Barger, A. J., \& Hu, E. M., Capak, P., \& Songaila, A.\ 2004b, \aj, 127, 3137

\reference{cram98}
Cram, L., Hopkins, A., Mobasher, B., \& Rowan-Robinson, M.\ 1998, \apj, 507, 155

\reference{croton06} 
Croton, D., Springel, V., White, S. D. M., et al.\ 2006, \mnras, 365, 11

\reference{dempsey13}
Dempsey, J. T., Friberg, P., Jenness, T., et al.\ 2013, \mnras, 430, 2534

\reference{dimatteo05} 
Di Matteo, T., Springel, V., \& Hernquist, L.\  2005, \nat, 433, 604

\reference{elbaz11}
Elbaz, D., Dickinson, M., Hwang, H. S., et al.\ 2011, A\&A, 533, 119

\reference{elvis09}
Elvis, M., Civano, F., Vignali, C., Puccetti, S., Fiore, F., et al.\ 2009, ApJS, 184, 158

\reference{faber03}
Faber, S. M., Phillips, A. C., Kibrick, R. I., et al.\ 2003, SPIE, 4841, 1657

\reference{farrah12}
Farrah, D., Urrutia, T., Lacy, M., et al.\ 2012, \apj, 745, 178

\reference{fazio04} 
Fazio, G. G., Hora, J. L., Allen, L. E., et al.\ 2004, \apjs, 154, 10

\reference{fritz06}
Fritz, J., Franceschini, A., \& Hatziminaoglou, E.\ 2006, \mnras, 366, 767

\reference{gandhi09}
Gandhi, P., Horst, H., Smette, A., et al.\ 2009, A\&A, 502, 457

\reference{giavalisco04} 
Giavalisco, M., Dickinson, M., Ferguson, H. C., et al.\ 2004, \apj, 600, L93

\reference{granato04} 
Granato, G. L., De Zotti, G., Silva, L., Bressan, A., \& Danese, L.\ 2004,
\apj, 600, 580

\reference{greve}
Greve, T. R., Pope, A., Scott, D., et al.\ 2008, \mnras, 389, 1489

\reference{griffin10} 
Griffin, M. J., Abergel, A., Abreu, A., et al.\ 2010, A\&A, 518, L3

\reference{grogin11} 
Grogin, N. A., Kocevski, D. D., Faber, S. M., et al.\ 2011, \apjs, 197, 35

\reference{harrison12} 
Harrison, C. M., Alexander, D. M., Mullaney, J. R.\ 2012, \apj, 760, L15

\reference{hasinger05} 
Hasinger, G., Miyaji, T., \& Schmidt, M.\ 2005, A\&A, 441, 417

\reference{hat10} 
Hatziminaoglou, E., Omont, A., Stevens, J. A., et al.\ 2010, A\&A, 518, L33

\reference{helou85} 
Helou, G., Soifer, B. T., \& Rowan-Robinson, M.\ 1985, \apj, 298, L7

\reference{holland13} 
Holland, W. S., Bintley, D., Chapin, E. L., et al.\ 2013, MNRAS, 430, 2513

\reference{ahopkins03}
Hopkins, A. M., Miller, C. J., Nichol, R. C., et al.\ 2003, \apj, 599, 971

\reference{hopkins06}
Hopkins, P. F., Hernquist, L., Cox, T. J., et al.\ 2006, \apjs, 163, 1

\reference{horst08}
Horst, H., Gandhi, P., Smette, A., \& Duschl, W. J.\ 2008, A\&A, 479, 389

\reference{hsieh12} 
Hsieh, B.-C., Wang, W.-H., Hsieh, C.-C., et al.\ 2012, \apjs, 203, 23

\reference{ibar10} 
Ibar, E., Ivison, R. J., Best, P. N., et al.\ 2010, \mnras, 401, L53

\reference{ichi12}
Ichikawa, K., Ueda, Y., Terashima, Y.\ 2012, \apj, 754, 45

\reference{ivison10}
Ivison, R. J., Alexander, D. M., Biggs, A. D., et al.\ 2010, \mnras, 402, 245

\reference{kennicutt98} 
Kennicutt, R. C.\ 1998, \apj, 498, 541

\reference{kimura10}
Kimura, M., Maihara, T., Iwamuro, F., et al.\ 2010, PASJ, 62, 1135

\reference{koekemoer11} 
Koekemoer, A. M., Faber, S. M., Ferguson, H. C., et al.\ 2011, \apjs, 197, 36

\reference{lafranca05} 
La Franca, F., Fiore, F., Comastri, A., et al.\ 2005, \apj, 635, 864

\reference{larson11} 
Larson, D., Dunkley, J., Hinshaw, G., et al.\ 2011, \apjs, 192, 16

\reference{lefevre05}
Le F{\`e}vre, O., Vettolani, G., Garilli, B., et al.\ 2005, A\&A, 439, 845

\reference{lehmer12} 
Lehmer, B. D., Xue, Y. Q., Brandt, W. N., et al.\ 2012, \apj, 752, 46

\reference{lilly07}
Lilly, S. J., Le F{\`e}vre, O., Renzini, A., et al.\ 2007, \apjs, 172, 70

\reference{lutz10} 
Lutz, D., Mainieri, V., Rafferty, D., et al.\ 2010, \apj, 712, 1287

\reference{lutz11} 
Lutz, D., Poglitsch, A., Altieri, B., et al.\ 2011, A\&A, 532, 90

\reference{magnelli13} 
Magnelli, B., Popesso, P., Berta, S., et al.\ 2013, A\&A, 553, 132

\reference{mark09} 
Markwardt, C.~B.\ 2009, in ASP Conf. Ser., 411, eds.~D.~A.~Bohlender,
D.~Durand, \& P.~Dowler, p.251

\reference{martin05} 
Martin, D. C., Fanson, J., Schiminovich, D., et al.\ 2005, \apj, 619, L1

\reference{mauch07} 
Mauch, T., \& Sadler, E. M.\ 2007, \mnras, 375, 931

\reference{mclean12} 
McLean, I. S., Steidel, C. C., Epps, H. W., et al.\ 2012, SPIE, 8446, 84460J

\reference{miller13} 
Miller, N. A., Bonzini, M., Fomalont, E. B., et al.\ 2013, \apjs, 205, 13

\reference{miyazaki02} 
Miyazaki, S., Komiyama, Y., Sekiguchi, M., et al.\ 2002, PASJ, 54, 833

\reference{moric10}
Mori{\'c}, I., Smol\v{c}i{\'c}, V., Kimball, A., Riechers, D. A., Ivezi{\'c}, \v{Z}., \& Scoville, N.\ 2010,
\apj, 724, 779

\reference{mullaney11}
Mullaney, J. R., Alexander, D. M., Goulding, A. D., \& Hickox, R. C.\ 2011, \mnras, 414, 1082

\reference{mullaney12} 
Mullaney, J. R., Pannella, M., Daddi, E., et al.\ 2012, \mnras, 419, 95

\reference{mush14} 
Mushotzky, R. F., Shimizu, T. T., Mel{\'e}ndez, M., \& Koss, M.\ 2014, \apj, 781, L34

\reference{netzer07} 
Netzer, H., Lutz, D., Schweitzer, M., et al.\ 2007, ApJ, 666, 806

\reference{nguyen10}
Nguyen, H. T., Schulz, B., Levenson, L., et al.\ 2010, A\&A, 518, L5

\reference{nonino09} 
Nonino, M., Dickinson, M., Rosati, P., et al.\ 2009, \apjs, 183, 244

\reference{oliver12}
Oliver, S. J., Bock, J., Altieri, B., et al.\ 2012, \mnras, 424, 1614

\reference{omont03}
Omont, A., Beelen, A., Bertoldi, F., et al.\ 2003, A\&A, 398, 857

\reference{orellana11}
Orellana, G., Nagar, N. J., Isaak, K. G., et al.\ 2011, A\&A, 531, A128

\reference{ostriker81}
Ostriker, J. P., \& Cowie, L. L.\ 1981, \apj, 243, L127

\reference{owen15}
Owen, F. N.\ 2015, \apj, in preparation

\reference{page12}
Page, M. J., Symeonidis, M., Vieira, J. D., et al.\ 2012, \nat, 485, 213

\reference{penner11} 
Penner, K., Pope, A., Chapin, E. L., et al.\ 2011, \mnras, 410, 2749

\reference{perera08}
Perera, T. A., Chapin, E. L., Austermann, J. .E., et al.\ 2008, \mnras, 391, 1227

\reference{pilbratt10}
Pilbratt, G. L., Riedinger, J. R., Passvogel, T., et al.\ 2010, A\&A, 518, L1

\reference{poglitsch10} 
Poglitsch, A., Waelkens, C., Geis, N., et al.\ 2010, A\&A, 518, L2

\reference{popesso09}
Popesso, P., Dickinson, M., Nonino, M., et al.\ 2009, A\&A, 494, 443

\reference{priddey03}
Priddey, R. S., Isaak, K. G., McMahon, R. G., \& Omont, A.\ 2003, \mnras, 339, 1183

\reference{rafferty11} 
Rafferty, D. A., Brandt, W. N., Alexander, D. M., et al.\ 2011, \apj, 742, 3

\reference{reddy06}
Reddy, N. A., Steidel, C. C., Erb, D. K., Shapley, A. E., \&
Pettini, M.\ 2006, \apj, 653, 1004

\reference{richards05}
Richards, G. T., Croom, S. M., Anderson, S. F., et al.\ 2005, \mnras, 360, 839

\reference{rieke04} 
Rieke, G. H., Young, E. T., Engelbracht, C.W., et al.\ 2004, \apjs, 154, 25

\reference{rosario12} 
Rosario, D. J., Santini, P., Lutz, D., et al.\ 2012, A\&A, 545, A45

\reference{rovilos12} 
Rovilos, E., Comastri, A., Gilli, R., et al.\ 2012, A\&A, 546, A58

\reference{sanders96} 
Sanders, D. B., \& Mirabel, I. F.\ 1996, ARA\&A, 34, 749

\reference{shao10} 
Shao, L., Lutz, D., Nordon, R., et al.\ 2010, A\&A, 518, L26

\reference{sijacki07}
Sijacki, D., Springel, V., Di Matteo, T., \& Hernquist, L.\ 2007, \mnras, 380, 877

\reference{silk98} 
Silk, J., \& Rees, M. J.\ 1998, A\&A, 331 L1

\reference{silverman14}
Silverman, J. D., Kashino, D., Arimoto, N., et al.\ 2015, \apjs, submitted (arXiv:1409.0447)

\reference{silverman10}
Silverman, J. D., Mainieri, V., Salvato, M., et al.\ 2010, \apjs, 191, 124

\reference{soifer08}
Soifer, B. T., Helou, G., \& Werner, M.\ 2008, ARA\&A, 46, 201

\reference{springel05} 
Springel, V., Di Matteo, T., \& Hernquist, L.\ 2005, \mnras, 361, 776

\reference{steffen03} 
Steffen, A. T., Barger, A. J., Cowie, L. L., Mushotzky, R. F., \& Yang, Y.\ 2003, \apj, 596, L23

\reference{szokoly04} 
Szokoly, G. P., Bergeron, J., Hasinger, G., et al.\ 2004, \apjs, 155, 271

\reference{thomson14}
Thomson, A. P., Ivison, R. J., Simpson, J. M., et al.\ 2014, \mnras, 442, 577

\reference{treister09}
Treister, E., Virani, S., Gawiser, E., et al.\ 2009, \apj, 693, 1713

\reference{trouille08} 
Trouille, L., Barger, A. J., Cowie, L. L., Yang, Y., \& Mushotzky, R. F.\ 2008, \apj, 179, 1

\reference{trump07}
Trump, J. R., Impey, C. D., McCarthy, P., et al.\ 2007, \apjs, 172, 383

\reference{ueda03} 
Ueda, Y., Akiyama, M., Ohta, K., \& Miyaji, T.\ 2003, \apj, 598, 886

\reference{wang13} 
Wang, S. X., Brandt, W. N., Luo, B., et al.\ 2013, \apj, 778, 179

\reference{wang10} 
Wang, W.-H., Cowie, L. L., Barger, A. J., Keenan, R. C., \&
Ting, H.-C.\ 2010, \apjs, 187, 251

\reference{wang11}
Wang, W.-H., Cowie, L. L., Barger, A. J., \& Williams, J. P.\ 2011, \apj, 726, L18

\reference{weiss09} 
Wei\ss, A., Kov{\'a}cs, A., Coppin, K., et al.\ 2009, \apj, 707, 1201

\reference{weisskopf02}
Weisskopf, M. C., Brinkman, B., Canizares, C., Garmire, G. P., Murray, S., \&
Van Speybroeck, L. P.\ 2002, PASP, 114, 1

\reference{wirth04}
Wirth, G. D., Willmer, C. N. A., Amico, P., et al.\ 2004, \aj, 127, 3121

\reference{xue10} 
Xue, Y. Q., Luo, B., Brandt, W. N., et al.\ 2011, \apjs, 195, 10

\end{references}
\end{document}